\documentclass[portuges,english]{iopart}
\usepackage{ae,aecompl}
\usepackage[T1]{fontenc}
\usepackage[latin9]{inputenc}
\usepackage{color}
\usepackage{babel}
\usepackage{units}
\usepackage{graphicx}
\usepackage[numbers]{natbib}
\usepackage[unicode=true,
 bookmarks=true,bookmarksnumbered=false,bookmarksopen=false,
 breaklinks=false,pdfborder={0 0 1},backref=false,colorlinks=false]
 {hyperref}

\makeatletter
\usepackage{iopams}
\usepackage{setstack}

\expandafter\let\csname overset\endcsname\relax
\expandafter\let\csname underset\endcsname\relax
\expandafter\let\csname sideset\endcsname\relax
\expandafter\let\csname uproot\endcsname\relax
\expandafter\let\csname leftroot\endcsname\relax
\expandafter\let\csname boxed\endcsname\relax
\expandafter\let\csname dddot\endcsname\relax
\expandafter\let\csname ddddot\endcsname\relax
\expandafter\let\csname substack\endcsname\relax
\expandafter\let\csname subarray\endcsname\relax
\expandafter\let\csname endsubarray\endcsname\relax
\expandafter\let\csname equation*\endcsname\relax
\expandafter\let\csname endequation*\endcsname\relax

\usepackage{amssymb}
\usepackage{amsmath}

\usepackage{babel}
\usepackage{babel}

\makeatother

\begin{document}
\selectlanguage{portuges}%
\global\long\def\ket#1{\left|#1\right\rangle }
 \global\long\def\bra#1{\left\langle #1\right|}
 \global\long\def\braket#1#2{\langle#1|#2\rangle}
 \global\long\def\inf{\infty}

\selectlanguage{english}%

\title{Spectral Functions of One-dimensional Systems with Correlated Disorder}

\author{N. A. Khan}

\address{Centro de Física das Universidades do Minho e Porto\\
 Departamento de Física e Astronomia, Faculdade de Ciências, Universidade
do Porto, 4169-007 Porto, Portugal}

\ead{nak@fc.up.pt}

\author{J. M. Viana Parente Lopes}

\address{Centro de Física das Universidades do Minho e Porto\\
 Departamento de Física e Astronomia, Faculdade de Ciências, Universidade
do Porto, 4169-007 Porto, Portugal}

\ead{jlopes@fc.up.pt}

\author{J. P. Santos Pires}

\address{Centro de Física das Universidades do Minho e Porto\\
 Departamento de Física e Astronomia, Faculdade de Ciências, Universidade
do Porto, 4169-007 Porto, Portugal}

\ead{up201201453@fc.up.pt}

\author{J. M. B. Lopes dos Santos}

\address{Centro de Física das Universidades do Minho e Porto\\
 Departamento de Física e Astronomia, Faculdade de Ciências, Universidade
do Porto, 4169-007 Porto, Portugal}

\ead{jlsantos@fc.up.pt}
\begin{abstract}
We investigate the spectral function of Bloch states in an one-dimensional
tight-binding non-interacting chain with two different models of static
correlated disorder, at zero temperature. We report numerical calculations
of the single-particle spectral function based on the Kernel Polynomial
Method, which has an $\mathcal{O}(N)$ computational complexity. These
results are then confirmed by analytical calculations, where precise
conditions were obtained for the appearance of a classical limit in
a single-band lattice system. Spatial correlations in the disordered
potential give rise to non-perturbative spectral functions shaped as the probability distribution of the random on-site energies, even at low disorder strengths. In the case of disordered potentials
with an algebraic power-spectrum, $\propto\left|k\right|^{-\alpha}$,
we show that the spectral function is not self-averaging for $\alpha\geq1$. 
\end{abstract}
\maketitle

\section{Introduction}

The one-electron spectral function is a key ingredient in the understanding
of interacting and of disordered electronic systems. It can be thought
of as the energy distribution of a state of momentum $\mathbf{k}$,
$\rho(\mathbf{k},E)$. In a non-interacting translationally invariant
system it is simply a Dirac delta function of energy, peaked at the
single particle energy $E_{\mathbf{k}}$.

The spectral function has been the subject of intense study in correlated
electronic systems, because it bears clear signatures of the low energy
phases of interacting electron systems, whether it be a Fermi Liquid\citep{Flensberg2004,Dzyaloshinski1975},
a marginal Fermi liquid as in high $T_{c}$ cuprates \citep{ISI:000087318700010},
or a one-dimensional Tomonaga-Luttinger liquid, with charge-spin separation
\citep{ISI:A1973Q362700043}. It is experimentally accessible by angle-resolved
photo-emission spectroscopy (ARPES) \citep{ISI:000182718200004,ISI:A1996VQ14700029}.

Random disorder can introduce a finite width on the spectral function,
averaged over disorder realizations, even in the absence of interactions.
The more common approaches to its calculation rely on the Born approximation
for the decay rate of a momentum state due to scattering by the disordered
potential. They implicitly (or explicitly) assume that the disordered
potential is a weak perturbation of the kinetic or band energy terms
of the Hamiltonian, and generally lead to a lorentzian line shape
for $\rho(\mathbf{k},E)$. One does not need strong disorder to ensure
localization in 1D or 2D, and in weak localization this approach to
the one particle spectral function is quite sufficient \citep{Muller2010}.

The concept of spectral function, however, is not confined to weak
disorder. Very efficient numerical methods are able to compute $\rho(\mathbf{k},E)$
for any strength of disorder \citep{Weibe2006}. Trappe, Delande and
Muller\citep{ISI:000355253100005} studied a continuum model with
correlated disorder and argued that when the root mean square of the
local random potential far exceeds the kinetic energy scale, $E_{\xi}=\hbar^{2}/2m\xi^{2}$
(where $\xi$ is the spatial correlation length of the disorder),
the non-commutativity of position and momentum can be ignored, and
a classical limit is achieved, in which the spectral function portrays
the probability distribution of the random potential. The coherent
potential approximation, a well-known approximation to treat disorder
problems \citep{ISI:A19679364600018,ISI:A1968C237800007}, which in
its original formulation cannot account for spatially correlated disorder,
has been generalized to treat spatially correlated disorder\citep{ISI:000271351500051}
and can also go beyond perturbation theory and reproduce the classical
limit results for strong disorder.

The exquisite control that has become available in ultra cold atom
experiments has renewed interest in the experimental study of disordered
potentials, free of the complication of interactions, always present
in electronic systems. Atomic clouds can be transferred into a random
potential created by laser speckle and several experiments have been
made on Anderson localization\citep{ISI:000269638200142,ISI:000295580300039,ISI:000303599200020},
included a measurement of the dependence of the mobility edge with
the strength of the disordered potential\citep{ISI:000357197300020}.

The random potential implemented in ultra-cold atom experiments is
correlated in space, in contrast with the standard Anderson model
of site disorder. Disorder correlation studies of Anderson localization
have been carried out for several decades now \citep{ISI:A1986D019900001}.
Significant results were obtained in 1D, where it was found that extended
states can exist at discrete energies in short-range correlated models
\citep{ISI:A1990DL89100023} and that a mobility edge appears in models
with power-law decay of spatial correlation of the random potential
\citep{ISI:000301813100001,ISI:000076616000046}.

Quite recently, a direct measurement of the one-particle spectral
function in an ultra-cold atom experiment was reported\citep{PhysRevLett.120.060404}.
By varying the intensity of the random potential, one observes a change
from a perturbative lorentzian shape, to an asymmetric line shape,
that reflects the probability distribution of the random potential.

Our focus in this paper is also on the spectral function in 1D tight-binding
models with correlated disorder. Unlike in the continuum case, band
models have an intrinsic kinetic energy scale given by the bandwidth.
It is relevant to consider whether the classical limit can be reached
even when disorder is weak, in the sense that the mean free path is
much larger than the lattice spacing. When the disorder correlation
length is much larger than the unit cell, scattering becomes \emph{local
in momentum-space} and we explore this feature to show analytically
how the classical limit emerges. Moreover, we also study the interesting
case of disorder correlations that decay as a power-law, with a characteristic
power spectrum $S\left(k\right)\sim1/k^{\alpha}$ \citep{ISI:000076616000046}.
This type of disorder has an infinite correlation length, and would
appear to be always in the classical limit. Instead, we find that
this limit for the averaged spectral function requires that $\alpha>1$,
when the scattering really becomes local in momentum space. Our results
are confirmed by numerical calculations.

Localization properties have been studied for these power-law spectrum
disorder models. While in the Anderson and other short-range correlated
models, all states are localized in 1D, for these power-law spectrum
models it has been claimed that a mobility edge appears for $\alpha\geq2$\citep{ISI:000076616000046}.
This conclusion has been contested, on the grounds that in the thermodynamic
limit this potential is not really disordered\citep{ISI:000319729900004}.
To investigate possible issues with the thermodynamic limit for these
models, we investigated the statistical properties of the spectral
function for different sized chains. We did find a transition from
self-averaging to non self-averaging behavior at $\alpha=1$. The
spectral function of even a very a large system will depend on the
specific realization of disorder it carries. It is significant, however,
that this transition occurs well below the value of $\alpha=2$. The
spectral function remains non self-averaging beyond $\alpha=2$, which
makes it hard to argue that the potential is not really disordered.

The rest of this paper is organized as follows. In the next section,
we start by defining our basic tight-binding model. Randomness is
introduced in the site energies, and is characterized by its Fourier
components, which have a prescribed magnitude, but randomly distributed
independent phases. We then briefly review the Kernel Polynomial Method
(KPM) as a tool for the numerical calculation of the spectral function.
In section III, we present our numerical results for $\rho\left(k,E\right)$,
and confirm our main findings by analytical calculations done in Section
IV. Additionally, some numerical results of the fluctuations of $\rho(k,E)$
and its self-averaging properties are also discussed. Finally, in
Section V we sum up our conclusions.

\section{The Disorder Model and The Kernel Polynomial Method}

\subsection{The Disorder Model }

The Hamiltonian we use is an one-dimensional tight-binding model with
nearest neighbor hopping and random site energies, 
\begin{align}
\mathcal{H}=\sum_{m=0}^{L-1}\varepsilon_{m}\ket{\varphi_{m}}\bra{\varphi_{m}}-t\left[\sum_{m=0}^{L-1}\ket{\varphi_{m+1}}\bra{\varphi_{m}}+\ket{\varphi_{m}}\bra{\varphi_{m+1}}\right]\label{eq:Hamiltonian}
\end{align}
where $\left\{ \ket{\varphi_{m}};\,m=0,\dots,L-1\right\} $ are the
local Wannier states. In what follows, we impose periodic boundary
conditions by setting $\ket{\varphi_{m}}=\ket{\varphi_{m+N}}$, the
lattice parameter $a$ is taken as $1$, and all energies are measured
in units of the hopping $t$ (\emph{i.e.,} $t=1$).

If there were no disorder, the exact eigenstates of the previous Hamiltonian
would be the Bloch states, defined as

\begin{equation}
\ket k=\frac{1}{\sqrt{L}}\sum_{m=0}^{L-1}e^{ikm}\ket{\varphi_{m}}.\label{eq:planwave-1-1}
\end{equation}
The presence of static disorder causes scattering of $\ket k\to\text{\ensuremath{\ket{k+q}}}$,
characterized by the matrix elements of the random potential $\mathcal{V}:=\sum_{m}\varepsilon_{m}\ket{\varphi_{m}}\bra{\varphi_{m}}$
that connect two Bloch states, \emph{i.e.}, 
\begin{equation}
\bra{k+q}\mathcal{V}\ket k=\frac{1}{L}\sum_{m}\varepsilon_{m}e^{-iqm},\label{eq:3}
\end{equation}
seen here to depend only on the transferred momentum $q$. We easily
invert Eq. \ref{eq:3} to express the local energies as the Fourier
sum 
\begin{equation}
\varepsilon_{m}=\sum_{q}\bra{k+q}\mathcal{V}\ket ke^{iqm}.\label{eq:corr pot}
\end{equation}
For the purposes of this paper, we choose to model the randomness
by taking these matrix elements as 
\begin{equation}
\bra{k+q}\mathcal{V}\ket k=V(q)e^{i\phi_{q}},\label{eq:k_kq_matrix_element}
\end{equation}
where $V(q):=\left|\bra{k+q}\mathcal{V}\ket k\right|$ is a specified
even function of $q$ and $\phi_{q}$ is a random phase with a uniform
probability distribution in the circle $\left[0,2\pi\right[$. The
different phases are independent variables except for the constraints
$\phi_{q}=-\phi_{-q}$, which ensure the hermiticity of the Hamiltonian.
With these definitions, the mean of the site energies is $\overline{\varepsilon_{m}}=\sum_{q}V(q)\overline{e^{i\phi_{q}}}e^{iqm}=V(0)$,
since the condition $\phi_{q}=-\phi_{-q}$ fixes $\phi_{0}=0$ and
the individual phase averages are zero otherwise, $\overline{e^{i\phi_{q}}}=\delta_{q,0}$.
As $\overline{\varepsilon_{m}}$ merely shifts the spectrum, we will
always choose $\overline{\varepsilon_{m}}=0$, meaning that $V(0)=0$.

In general, the values of the energies in different sites will be
correlated in this model of disorder. The two-site covariance of the
potential can be written as 
\begin{equation}
\overline{\varepsilon_{n}\varepsilon_{m}}=\sum_{q,q'\neq0}V(q)V(q')\overline{e^{i\phi_{q'}}e^{i\phi_{q}}}e^{i\left(qm+q'n\right)},\label{eq:6}
\end{equation}
where all the phase averages factorize (unless $q=-q'$) and the average
of a single phase is zero,

\begin{subequations}

\begin{align}
\overline{e^{i\phi_{q}}} & =0,\\
\overline{e^{i\phi_{q'}}e^{i\phi_{q}}} & =\delta_{q+q',0}.
\end{align}
\end{subequations}Hence (using the property $V(q)=V(-q)$) 
\begin{equation}
\overline{\varepsilon_{n}\varepsilon_{m}}=2\sum_{q>0}V^{2}(q)\cos\left(q(n-m)\right).\label{eq:disorder_corr}
\end{equation}

From Eq. \ref{eq:disorder_corr}, we see that $V^{2}(q)$ can be related
to the Fourier transform of the spatial correlation function $C\left(n\right)$
of the disordered potential, as follows 
\begin{equation}
V^{2}(q):=\frac{1}{L}\sum_{n}\overline{\varepsilon_{n}\varepsilon_{0}}e^{iqn}=\frac{1}{L}\sum_{n}C\left(n\right)e^{iqn}.
\end{equation}
In the case of an uncorrelated disorder, like in the usual Anderson's
model, we have 
\begin{equation}
\overline{\varepsilon_{n}\varepsilon_{m}}=\sigma_{\varepsilon}^{2}\delta_{n,m},
\end{equation}
with $\sigma_{\varepsilon}^{2}:=\overline{\varepsilon^{2}}$, or,
equivalently 
\begin{equation}
V^{2}\left(q\right)=\frac{1}{L}\sigma_{\varepsilon}^{2}.
\end{equation}
Thus, for these models, the magnitude of the scattering matrix element
from $k\to k+q$ is independent of the transferred momentum, $q$
.

\subsubsection{Gaussian Correlated Disorder}

Our first model of correlated disorder is the gaussian case. For that,
we choose 
\begin{equation}
V(q):=\frac{A(q_{c})}{\sqrt{L}}\exp\left(-q^{2}/4q_{c}^{2}\right),\label{eq:Gaussian Potential}
\end{equation}
where $A(q_{c})$ is a measure for the strength of disorder. The $L^{-\frac{1}{2}}$
factor in Eq.\ref{eq:Gaussian Potential} is introduced in order to
have a well-defined thermodynamic limit for the local variance and
correlation functions of the disordered potential.

In this model, the values of $V(q)$ are only significant inside an
interval of linear size $q_{c}$, centered around $q=0$. This means
that the disordered potential couples Bloch states with nearby momenta,
more strongly\footnote{However, this does not mean an absence of back-scattering, since the
full effect of this potential must take all the multiple scattering
processes into account. As a matter of fact, these disordered potentials
with short-range correlations are believed to cause an exponential
localization of the eigenstates, in a manner similar to the one-dimensional
Anderson model with uncorrelated disorder.}. The statistical properties of the corresponding potential in the
$L\rightarrow\infty$ limit, can be calculated through Eq.~\ref{eq:disorder_corr},
yielding\begin{subequations} 
\begin{align}
\overline{\varepsilon^{2}}=\sigma_{\epsilon}^{2}= & A^{2}(q_{c})\int_{-\pi}^{\pi}\frac{dq}{2\pi}e^{-q^{2}/2q_{c}^{2}},\label{eq:gaussian variance}\\
\overline{\varepsilon_{n}\varepsilon_{m}}=A^{2}(q_{c}) & \int_{-\pi}^{\pi}\frac{dq}{2\pi}e^{-q^{2}/2q_{c}^{2}}e^{iq(n-m)}.
\end{align}
\end{subequations}

From these two equations, we notice that the normalized correlation
function does not depend on the parameter $A(q_{c})$, i.e. 
\begin{equation}
\Gamma(n-m):=\frac{\overline{\varepsilon_{n}\varepsilon_{m}}}{\sigma_{\varepsilon}^{2}}=\frac{\int_{-\pi}^{\pi}\frac{dq}{2\pi}e^{-q^{2}/2q_{c}^{2}}e^{iq(n-m)}}{\int_{-\pi}^{\pi}\frac{dq}{2\pi}e^{-q^{2}/2q_{c}^{2}}}.
\end{equation}
Finally, all the integrals above can be done analytically in the limit
when $q_{c}\ll\pi$. In this case, the integration intervals may be
extended to $k\in\left]-\infty,+\infty\right[$ and we get, 
\begin{equation}
\Gamma(m)=\exp\left(-\frac{q_{c}^{2}m^{2}}{2}\right).
\end{equation}
The correlation function of site energies is gaussian in real space
with a decay length $\xi=q_{c}^{-1}$. In this same limit, we can
also relate the parameter $A(q_{c})$ with the local disorder strength
using Eq. \ref{eq:gaussian variance}, i.e. 
\begin{equation}
A^{2}(q_{c})=\sqrt{2\pi}\frac{\sigma_{\varepsilon}^{2}}{q_{c}},
\end{equation}
meaning that, 
\begin{equation}
V(q)=\left(2\pi\right)^{\frac{1}{4}}\frac{\sigma_{\varepsilon}}{\sqrt{q_{c}L}}\exp\left(-q^{2}/4q_{c}^{2}\right).
\end{equation}

\subsubsection{Power-Law Correlated Disorder}

For our second model of disorder, we take the power-law potential
defined by De Moura and Lyra\citep{ISI:000076616000046}, for a periodic
chain of $L$ sites, as

\begin{equation}
\varepsilon_{m}=2A(\alpha)\sum_{p=1}^{L/2}\left(\frac{2\pi}{L}\right)^{\frac{1-\alpha}{2}}\frac{1}{p^{\frac{\alpha}{2}}}\cos\left(\frac{2\pi mp}{L}+\phi_{p}\right).\label{eq:power-law-disorder}
\end{equation}
The phases $\phi_{p}$ have the same properties as before, being uniformly
distributed in $\left[0,2\pi\right[$. We can reduce this definition
to our formulation by writing the Bloch wave-numbers as 
\begin{equation}
q:=\frac{2\pi}{L}p,
\end{equation}
so that Eq.~\ref{eq:power-law-disorder} becomes 
\begin{equation}
\varepsilon_{m}=2A(\alpha)\left(\frac{2\pi}{L}\right)^{\frac{1}{2}}\sum_{q>0}\frac{1}{q^{\alpha/2}}\cos\left(qm+\phi_{q}\right).
\end{equation}
Since this sum is carried only over the positive half of the first
Brillouin zone (i.e., $q=2\pi p/L$, $p=1,\dots L/2$), it can be
rewritten as 
\begin{align}
\varepsilon_{m} & =\sum_{q\neq0}V(q)e^{i\phi_{q}}e^{iqm},\label{eq:site_energy}
\end{align}
with $V(q)$ defined as 
\begin{equation}
V(q)=A(\alpha)\left(\frac{2\pi}{L}\right)^{\frac{1}{2}}\frac{1}{\left|q\right|^{\frac{\alpha}{2}}},\label{eq:power_law_Vq}
\end{equation}
and the independent random phases obeying the constraint $\phi_{q}=-\phi_{-q}$.
The $q=0$ term is excluded as before, and we have introduced a normalization
factor $A(\alpha)$ that will define a finite variance for the local
disorder.

To study the thermodynamic limit ($L\to\infty$) in the previous case
(gaussian), we replaced all the sums over $q$ by integrals. In this
case, since $q=2\pi p/L$, $p\in\mathbb{Z}\setminus\left\{ 0\right\} ,$
we could try to do the same, but this turns out to be quite tricky
due to the possibility of generating low-$q$ singularities. Consider,
as an example, the calculation of the disorder's local variance, 
\begin{equation}
\sigma_{\varepsilon}^{2}=\sum_{q\neq0}V^{2}\left(q\right)=A^{2}(\alpha)\left(\frac{2\pi}{L}\right)\sum_{q\neq0}\frac{1}{\left|q\right|^{\alpha}},\label{eq:24}
\end{equation}
for $\alpha<1,$ the corresponding integral does not have a low-$q$
singularity and the situation is be very similar to a system with
uncorrelated disorder. A more interesting case happens for $\alpha>1$,
where the integrals will have low-$q$ singularities with a natural
cut-off of $2\pi/L$. At the same time, in this case, the corresponding
sum over $p$ in 
\begin{equation}
\sigma_{\varepsilon}^{2}=2A^{2}(\alpha)\left(\frac{2\pi}{L}\right)^{1-\alpha}\sum_{p=1}^{L/2}\frac{1}{p^{\alpha}},\label{eq:sum}
\end{equation}
is found to converge as $L\to\infty$. These two facts mean that,
no matter how large $L$ is, the number of terms contributing to the
sum is always of $\mathcal{O}(1)$. Hence, we can never approximate
it by an integral. Luckily, the infinite sum in Eq.$\;$\ref{eq:sum}
is known to define the Riemann Zeta function\citep{ZetaFuntion},
\begin{equation}
\sum_{p=1}^{\infty}\frac{1}{p^{\alpha}}:=\mathcal{\zeta}(\alpha).
\end{equation}
Finally, in the same limit, the local variance of the disorder can
be written as 
\begin{equation}
\sigma_{\varepsilon}^{2}=2A^{2}(\alpha)\left(\frac{2\pi}{L}\right)^{1-\alpha}\zeta(\alpha),
\end{equation}
allowing us to express $A^{2}(\alpha)$ in terms of $\sigma_{\varepsilon}$,
as follows, 
\begin{equation}
A^{2}(\alpha)=\frac{\sigma_{\varepsilon}^{2}}{2\mathcal{\zeta}(\alpha)}\left(\frac{2\pi}{L}\right)^{\alpha-1}.\label{eq:norm_constant}
\end{equation}
The correlation function of this potential can also be calculated
using, 
\begin{align*}
\overline{\varepsilon_{n}\varepsilon_{m}} & =\sum_{q\neq0}V^{2}(q)e^{iq\left(n-m\right)}=\frac{4\pi A^{2}(\alpha)}{L}\sum_{q>0}\frac{1}{\left|q\right|^{\alpha}}\cos\left(q(n-m)\right).
\end{align*}
Writing $q=2\pi p/L$ and taking the thermodynamic limit in the last
sum, we can express the result in terms of a polylogarithm function
\citep{ZetaFuntion,ISI:000319729900004}, $Li_{\alpha}(z):=\sum_{p=1}^{\infty}z^{p}/p^{\alpha}$,
as follows, 
\begin{equation}
\Gamma_{\alpha}(m)=\frac{\overline{\varepsilon_{0}\varepsilon_{m}}}{\sigma_{\varepsilon}^{2}}=\frac{1}{\mathcal{\zeta}(\alpha)}Re\left[Li_{\alpha}\left(e^{-\frac{2\pi im}{L}}\right)\right].
\end{equation}

A plot of this space correlation function is shown in the Figure$\;$\ref{fig:polilog},
for several values of the exponent $\alpha$\textcolor{red}{\citep{ISI:000319729900004}}. 
\begin{center}
\begin{figure}[h]
\centering{}\includegraphics[width=10cm,height=6.5cm]{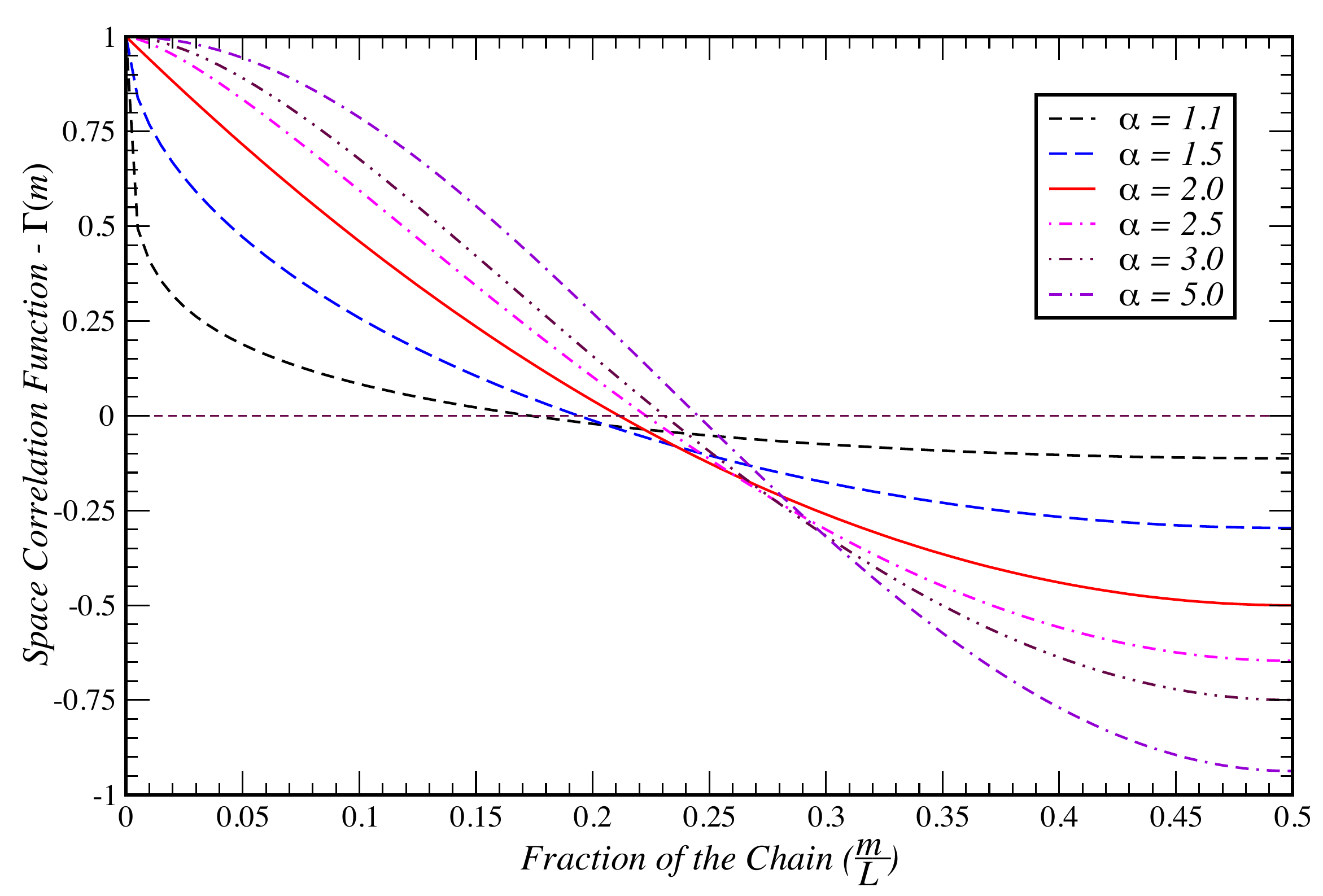}\caption{Plot of the space correlation for the power-law disordered potential
as a function, calculated for several values of the exponent $\alpha$.
The $\alpha\rightarrow+\protect\inf$ limit yields a perfect cosine
function, corresponding to an ordered system with an applied modulated
potential.\label{fig:polilog}\textcolor{red}{\citep{ISI:000319729900004}}}
\end{figure}
\par\end{center}

As a last remark, we note that to ensure a finite local variance,
$\sigma_{\varepsilon}$, we had to choose $A^{2}(\alpha)\propto1/L^{\alpha-1}$
(see Eq.~\ref{eq:norm_constant}). This weird fact implies that $A^{2}(\alpha)\to0$,
as $L\to\infty$ (for $\alpha>1$), which will have important consequences
in what follows.

\subsection{The Kernel Polynomial Method}

The spectral function of a large disordered quantum system can be
efficiently computed by a polynomial expansion-based technique ---
the\textbf{ Kernel Polynomial Method} (KPM) \citep{Silver1996,Roder1997,Weibe2006,Lin2016,jvlops2016}.
In this approach, a function of an operator with spectrum normalized
to the interval $\left]-1,1\right[$ is approximated by a truncated
Chebyshev series. The expansion coefficients can be computed either
by the stochastic evaluation of a trace \citep{Weibe2006,Lin2016}
or by the expectation values of Chebyshev polynomials in a given basis.
Furthermore, the accuracy and numerical convergence of the KPM estimates
are controlled by employing an optimized Gibbs damping factor and
using sufficient number of Chebyshev polynomials \citep{Weibe2006}.
The Chebyshev polynomial of the first kind, $T_{n}(x)$, is an $n^{th}$-degree
polynomial in $x$, defined as

\begin{equation}
T_{n}(x)=\cos(n\,\arccos(x)),\quad n\in\mathbb{N}\label{eq:polynomial}
\end{equation}
where $x$ takes values in the interval $]-1,1[$. Moreover, the $T_{n}(x)$'s
are generated by the following recurrence relations\begin{subequations}
\begin{align}
T_{0}(x) & =1,\,\,\,\,\,\,T_{1}(x)=x,\label{eq:Recursion}\\
T_{n+1}(x) & =2xT_{n}(x)-T_{n-1}(x),\label{eq:Recursion1}
\end{align}
\end{subequations} and also satisfy the orthogonality relation 
\begin{equation}
\int_{-1}^{1}\,T_{n}(x)T_{m}(x)(1-x^{2})^{-1/2}dx=\frac{\pi}{2}\delta_{n,m}(\delta_{n,0}+1).\label{eq:Orthognal}
\end{equation}

In our case, we consider a free electron gas hopping on a finite cyclic
chain of size $L$, under the influence of on-site correlated disorder.
Suppose that the $L\times L$ Hamiltonian matrix $\mathcal{H}$ (Eq.~\ref{eq:Hamiltonian}),
has eigenvalues $E_{\beta}$ with corresponding eigenstates $\left|\Psi_{\beta}\right\rangle $.
Then its zero temperature spectral function has the form 
\begin{equation}
\rho(k,E)=\sum_{\beta=0}^{L-1}\left|\left\langle k|\Psi_{\beta}\right\rangle \right|^{2}\delta(E-E_{\beta}),\label{eq:SDOS}
\end{equation}
where $\ket k$ is a Bloch state of one electron as defined in last
section. Notice also that, in the absence of disorder $\rho(k,E)=\delta(E-E_{k})$,
and by summing $\rho(k,E)$ over $k$ one obtains the density of states.

To calculate $\rho(k,E)$ we must normalize the Hamiltonian, so that
its spectrum fits inside the interval $]-1,1[$\footnote{The Hamiltonian and all energy parameters are rescaled by dividing
by $(2Dt+F)$, where $D$ is the dimension of the hypercubic lattice
system, $t$ the hopping, and $F$ is a number chosen so that in all
cases the spectrum of the Hamiltonian fits into the interval $]-1,1[$. }. The KPM approximation to the spectral function is written as 
\begin{equation}
\rho_{M}(k,E)=\frac{2}{\pi\sqrt{1-E^{2}}}\sum_{n=0}^{M-1}\frac{g_{n}\mu_{n}}{(1+\delta_{n,0})}T_{n}(E),\label{eq:kpm}
\end{equation}
where the expansion coefficients $\mu_{n}$ are determined as 
\begin{equation}
\begin{array}{ccc}
\mu_{n} & = & \int_{-1}^{1}T_{n}(E)\rho(k,E)\,dE=\bra kT_{n}(\mathcal{H})\ket k.\end{array}\label{eq:ChebM}
\end{equation}
The recursion relations obeyed by the Chebyshev polynomials carry
over to these moments, and greatly simplify their calculation. The
expression Eq.~\ref{eq:kpm}, represents the truncated sum of the
Chebyshev series. It is known that the abrupt truncation of the series
introduces Gibbs oscillations in the function to be approximated.
This phenomenon can be filtered out by employing an optimized damping
factor. The most appropriate and the one that we use here is the so-called
\textbf{Jackson Kernel} $g_{n}$ \citep{Silver1996} defined as follows
\begin{equation}
g_{n}=\frac{(M-n+1)\cos(\frac{n\pi}{M+1})}{M+1}+\frac{\sin(\frac{n\pi}{M+1})\cot(\frac{\pi}{M+1})}{M+1}.\label{eq:Jackson}
\end{equation}
The use of this kernel does not alter the series' convergence to the
intended function, as $M$ goes to infinity. Furthermore, this makes
the KPM approximations always non-negative, which is particularly
relevant when approximating a non-negative function, like $\rho(k,E)$.

\section{Numerical Results and Discussion}

We have performed numerical computations of the spectral functions
for the 1D non-interacting system in the presence of on-site gaussian
and power-law correlated disorder with periodic boundary conditions,
at zero temperature. The computations were carried out by using the
KPM. For comparison, we also include some results for the usual Anderson
Model.

\subsection{Gaussian Correlated Disorder}

We start by presenting results for the spectral function in the uncorrelated
Anderson model. For a rectangular distribution of site energies, 
\begin{equation}
P(\varepsilon_{n})=\frac{1}{W}\Theta\left(\frac{W}{2}-\varepsilon_{n}\right),
\end{equation}
and $\sigma_{\varepsilon}^{2}=\nicefrac{W^{2}}{12}$. The strength
of disorder is commonly characterized by $W$, but as we are interested
in other types of distributions for the site energies, in this paper
we use $\sigma_{\varepsilon}$ instead.

\begin{figure}[h]
\begin{centering}
\includegraphics[width=10cm,height=6.5cm]{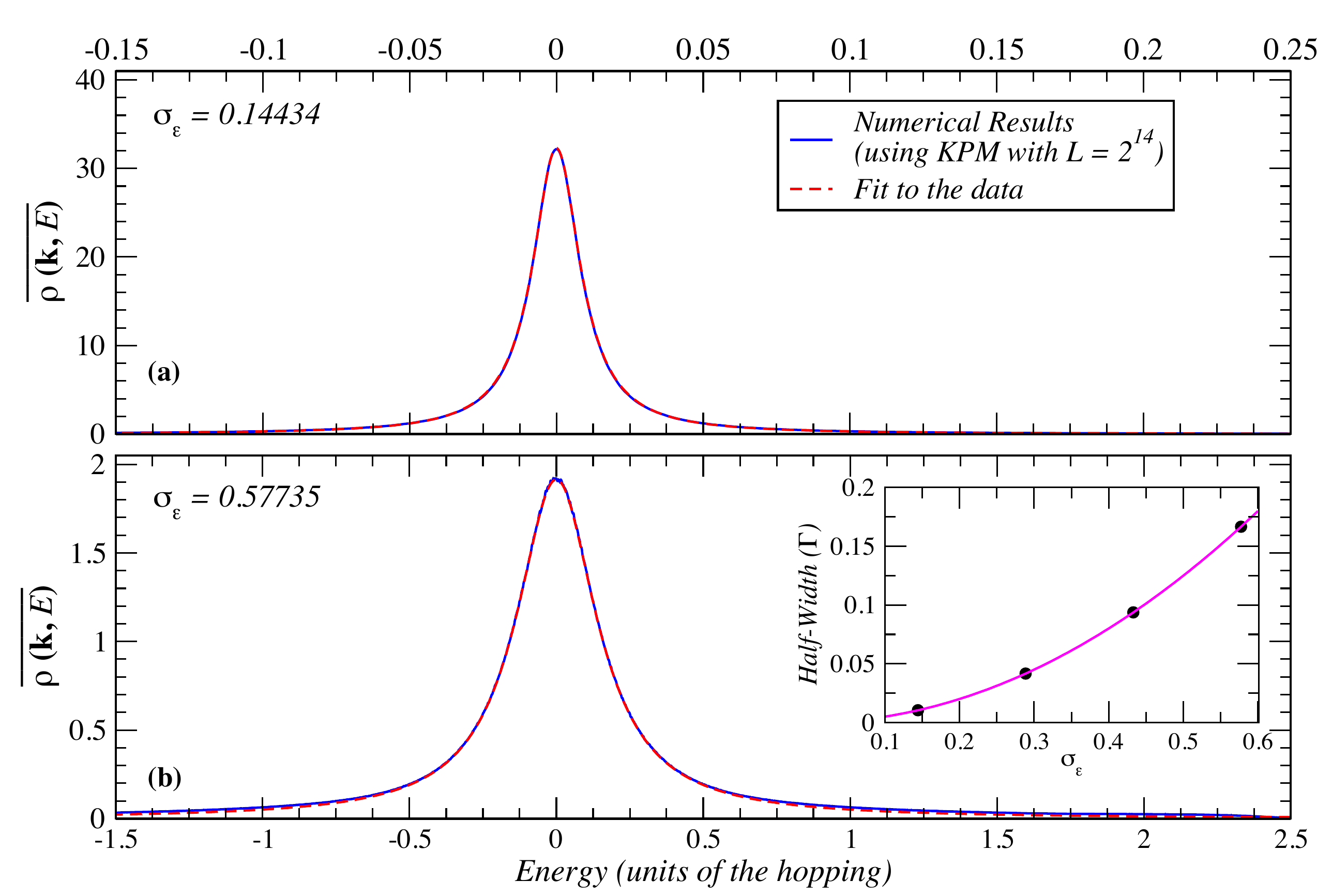} 
\par\end{centering}
\caption{The disorder-averaged spectral function $\rho(k,E)$ of the Anderson
model at the band center ($k=\pi/2,$ $E_{k}=0$) for different local
variances of the uncorrelated disorder $\sigma_{\varepsilon}^{2}$.
The spectral function is well represented by a lorentzian, as expected
for low disorder. The black dots in the inset are the corresponding
half-widths of the fitted curves; the magenta line is the Born approximation,
Eq.~\ref{eq:BornApp}.\label{fig:Spectral-function-of_Anderson}}
\end{figure}

In Figure~\ref{fig:Spectral-function-of_Anderson} we show the approximated
spectral function for various values of the local variance $\sigma_{\varepsilon}^{2}$,
at the band center, i.e. $E_{k}=0$ $(k=\pi/2)$. The data is well
fitted by a lorentzian, as expected from perturbation theory. In the
inset, we show a comparison between the half-width of the lorentzian,
obtained from the fits, and the value calculated from the Born approximation.
\begin{equation}
\hbar\Gamma=\frac{\sigma_{\varepsilon}^{2}}{2}.\label{eq:BornApp}
\end{equation}
This perturbative result seems to give a good account of the data
until values $\sigma_{\varepsilon}\lesssim1$.

\begin{figure}[h]
\begin{centering}
\includegraphics[width=6.45cm,height=5.6cm]{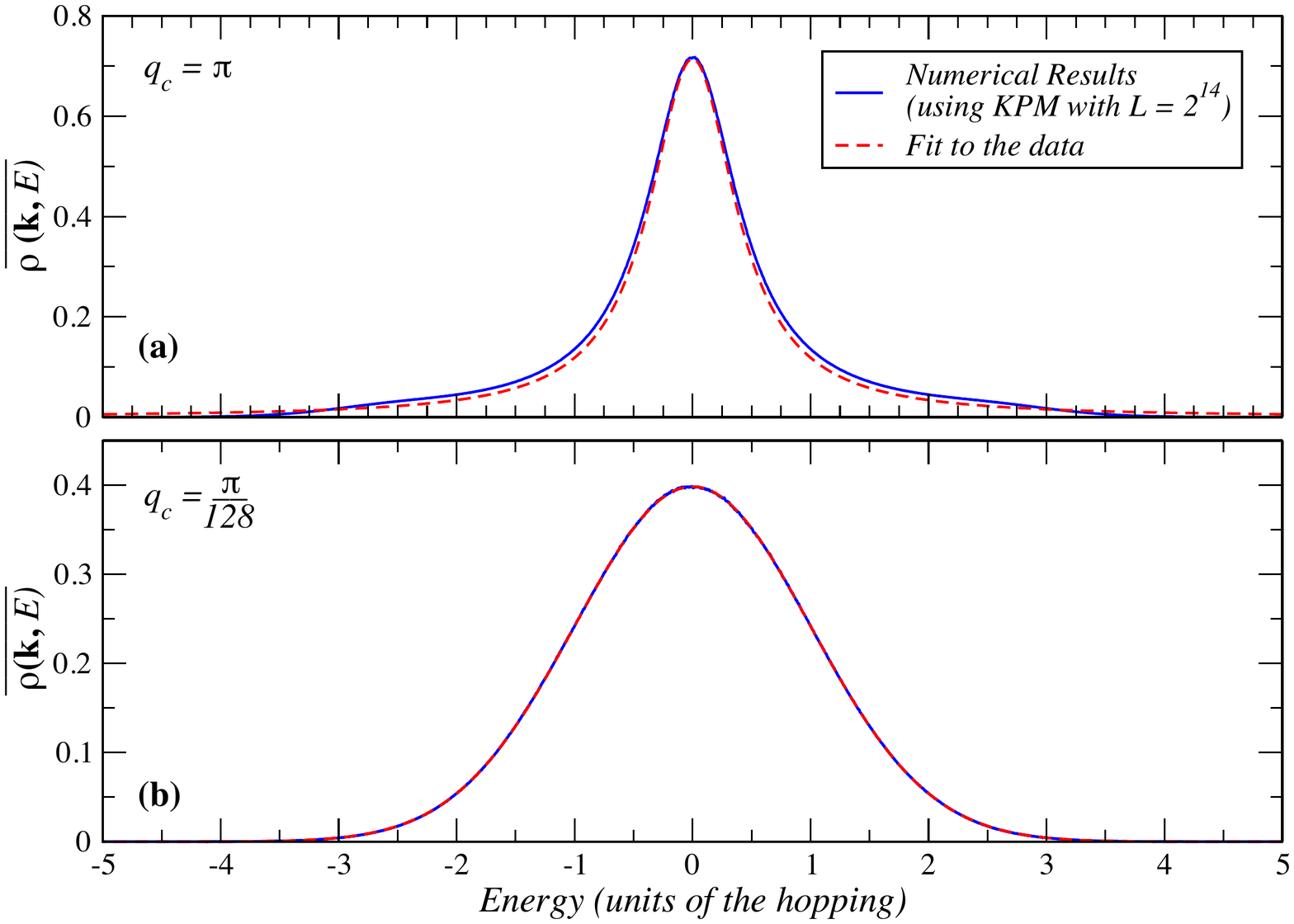}\includegraphics[width=6.5cm,height=5.6cm]{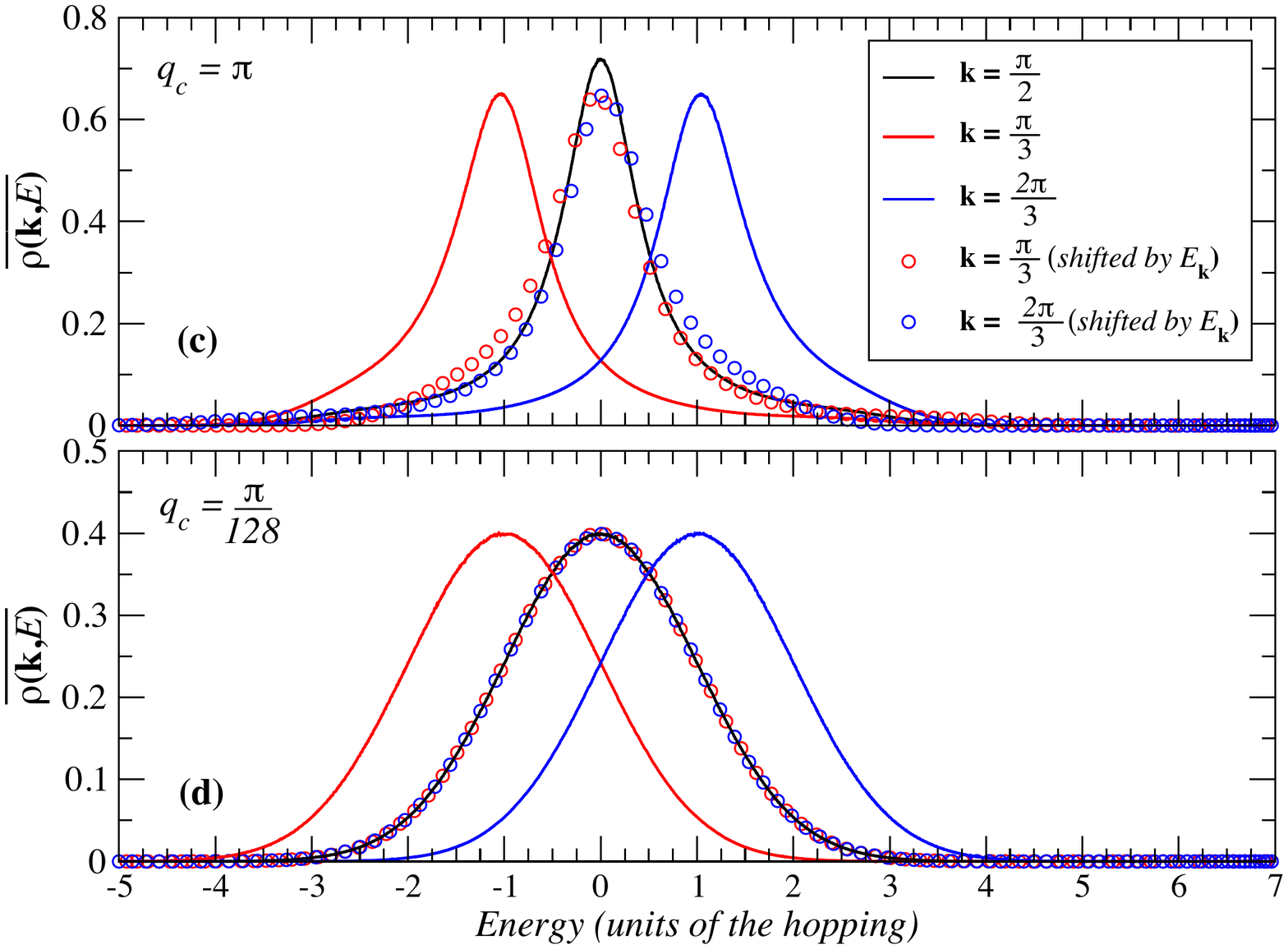} 
\par\end{centering}
\caption{The disorder-averaged spectral function for gaussian correlated disorder
with unit variance $\sigma_{\epsilon}^{2}=1,$ for (a) $q_{c}=\pi$,
and (b) $q_{c}=\pi/128.$ The spectral functions are reasonably fitted
by a lorentzian (upper panel) of half-width $\Gamma\backsimeq0.4456$,
and very well fitted a gaussian (lower panel) of variance $\sigma_{\varepsilon}^{2}=1$.
The plots (c) and (d), show the spectral function for different values
of $k$ and how it relates to its shape at the band's center. \label{fig:Gaussian_SF} }
\end{figure}

The spectral function, at the band center ($E_{k}=0),$ for a gaussian
correlated disorder with different values of the parameter $q_{c}$,
is shown in Figure~\ref{fig:Gaussian_SF} for $\sigma_{\varepsilon}=1$.
The magenta dashed curves are the corresponding fits. For $q_{c}=\pi$
{[}Figure~\ref{fig:Gaussian_SF}(a){]}, the best fit of the numerical
data can be found with a lorentzian of width $\Gamma\backsimeq0.4456$.

\begin{figure}[h]
\begin{centering}
\includegraphics[width=10cm,height=6.5cm]{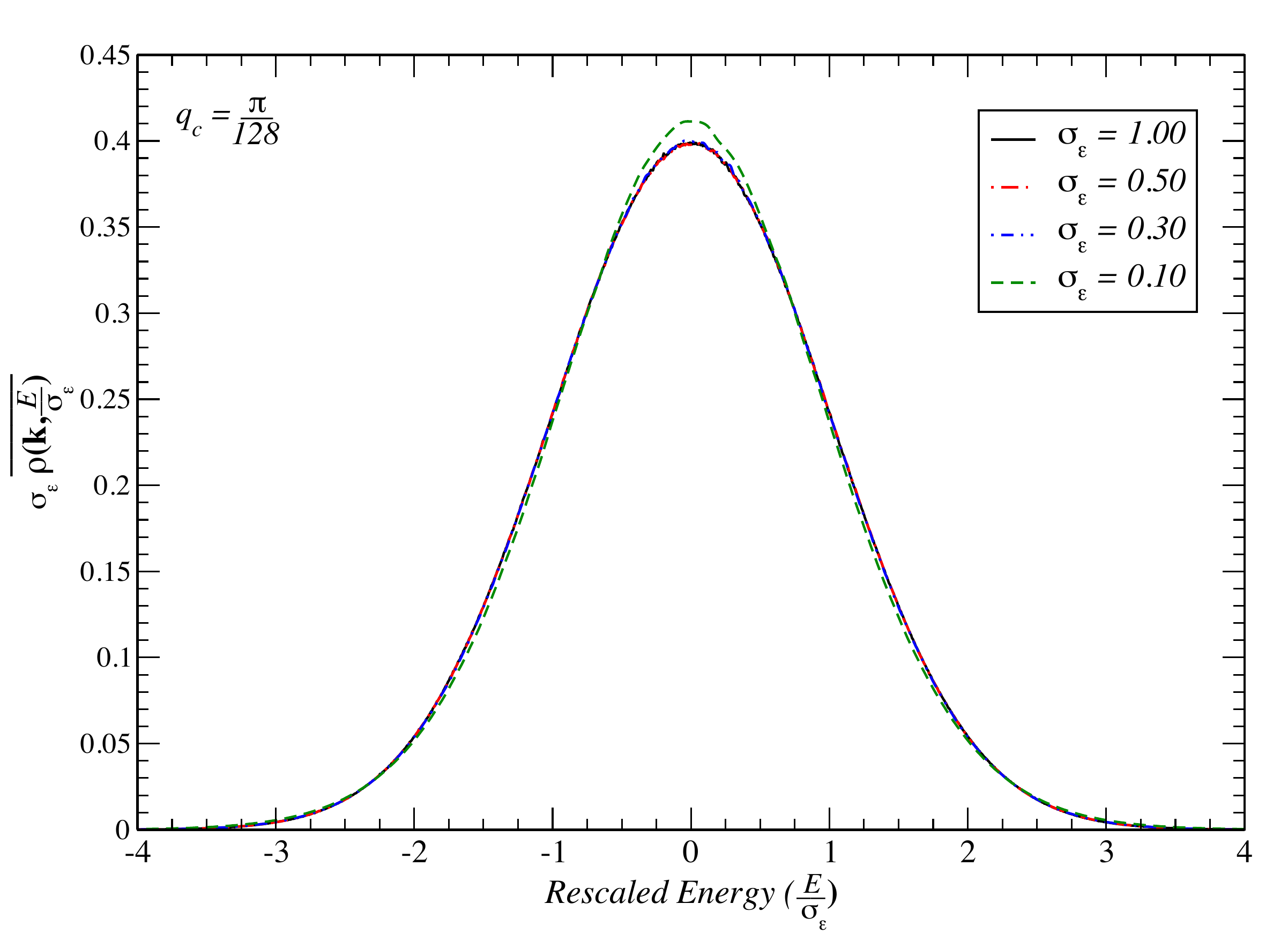} 
\par\end{centering}
\caption{The normalized spectral function for the gaussian correlated disorder
of the system of size $L=2^{14}$ with $8192$ Chebyshev coefficients
for different values of disorder variance $\sigma_{\varepsilon}$
.\label{fig:Scaled_Gaussian-1} }
\end{figure}

When $q_{c}=\pi/128$, the scattering becomes local in momentum space,
and the spectral function is seen to be a gaussian {[}Figure~\ref{fig:Gaussian_SF}(b){]}.
Its width is just the variance of the site energies, $\sigma_{\varepsilon}^{2},$
as can be seen in Figure~\ref{fig:Scaled_Gaussian-1}, where the
spectral functions for different values of $\sigma_{\varepsilon}$
are scaled to show that

\textbf{ 
\begin{equation}
\overline{\rho(k=\pm\frac{\pi}{2},E)}=\sigma_{\varepsilon}^{-1}\mathcal{N}\left(0,1,E/\sigma_{\varepsilon}\right),\quad\textrm{for }\sigma_{\varepsilon}>>\hbar v_{k}q_{c}.\label{eq:scalinggaussian}
\end{equation}
}In Eq. \ref{eq:scalinggaussian}, $\mathcal{N}(\mu,\sigma,\varepsilon)$
is the normal distribution of mean $\mu$ and variance $\sigma$.

This result calls to mind the classical limit of the spectral function
discussed by Trappe \emph{et. al.} \citep{ISI:000355253100005}. In
that limit, the disordered potential dominates, and the spectral function
merely reflects the probability distribution of local potential values.
This is, in fact, what is observed here. Since 
\begin{equation}
\varepsilon_{m}=2\sum_{q>0}V(q)\cos\left(qm+\phi_{q}\right),
\end{equation}
in the thermodynamic limit (i.e. $q_{c}\gg\pi/L$), the energy at
each site is a sum of a large number of random independent variables,
and by the central limit theorem, it is normally distributed. But
what is significant here is that this limit can be obtained even when
the disorder strength is small enough to be considered a weak perturbation
when compared to the bandwidth. As we will see later this will turn
out to be a consequence of the local character of the scattering in
momentum space.

\subsection{Power-Law Correlated Disorder}

\begin{figure}[h]
\begin{centering}
\includegraphics[width=6.35cm,height=5.7cm]{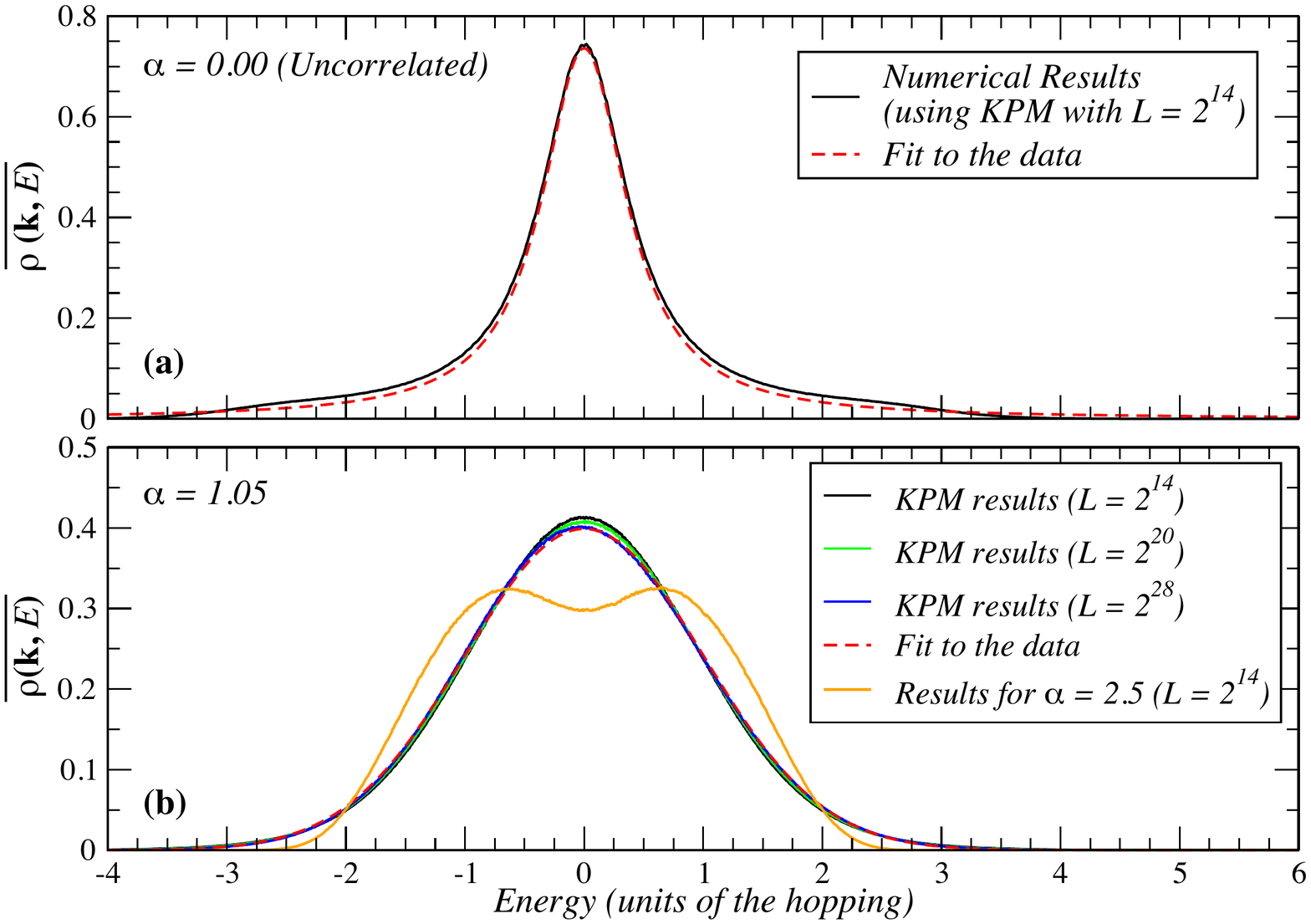}\includegraphics[width=6.5cm,height=5.83cm]{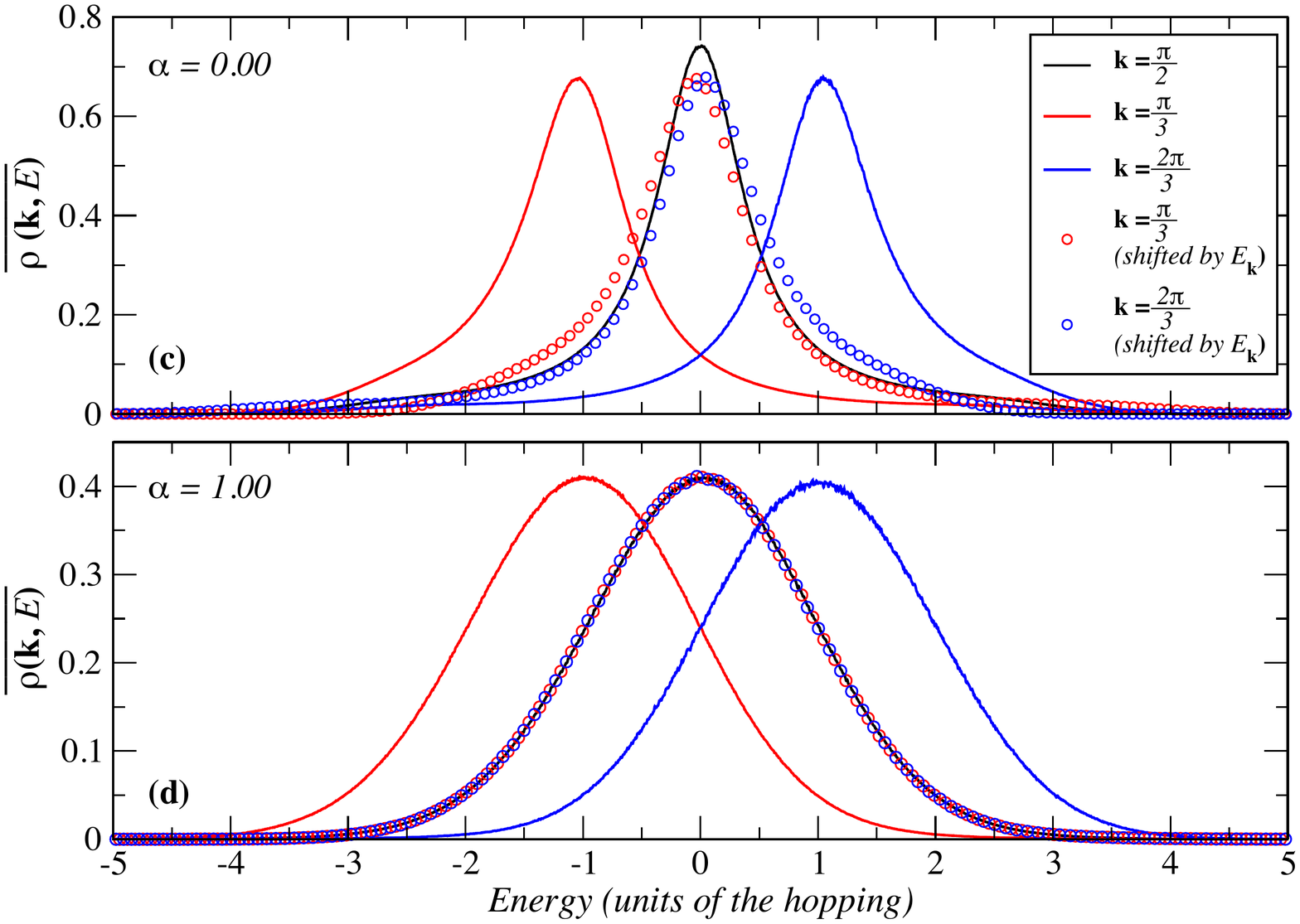} 
\par\end{centering}
\caption{Mean spectral function for two values of the correlation exponent,
$\alpha$, and $\sigma_{\varepsilon}=1.$ The numerical plots show
a good agreement with the following conclusions: (a) and (c) a lorentzian
fit of width $\Gamma\thicksim0.433$ and a strong dependence of the
shape with the value of $k$, for $\alpha<1$; (b) and (d) a gaussian
fit of unit variance in the large $L$ limit and a very weak dependence
of the shape with the value of $k$ for $\alpha\sim1^{+}$. In (b)
an example of numerical data for $\alpha=2.5$ illustrates the double-peak
structure which emerges for higher values of the exponent.\label{fig:PowerLaw-SF} }
\end{figure}

A power-law correlated disorder is characterized by the exponent $\alpha$
that determines how fast the Fourier transform of $\varepsilon_{n}$
decays with the wavenumber $q$,

\begin{equation}
V^{2}(q)\sim\frac{1}{\left|q\right|^{\alpha}}.
\end{equation}
As $\alpha$ increases, scattering becomes increasingly dominated
by small values of $q$ ($q\ll\pi)$. In Figure~\ref{fig:PowerLaw-SF},
we see that a transition for a lorentzian to a gaussian shape (with
unit variance) of the spectral function at the band center and for
$\sigma_{\varepsilon}=1$, occurs at $\alpha\approx1.0$. This transition
seems to hold for other values of $k$ as well, as can be seen from
the left panels of Figure \ref{fig:PowerLaw-SF}.\footnote{The shape of the lorentzian $\overline{\rho(k,E)}$ depends much more
strongly on the value of of $k$. This can be understood as the combined
effect of a change in the central velocity (which affects the mean
free path, i.e. the width) and the fact that the algebraic tails start
to feel the effect of the finite bandwidth.} On closer scrutiny, however, a perfect gaussian fit is only possible
for $\alpha\to1^{+}$, in the large $L$ limit, and deviations become
increasingly obvious as $\alpha$ increases; the spectral function
develops a two peaked structure as a function of energy, as shown
in Figure \ref{fig:PowerLaw-SF}(b) in orange.

\begin{figure}[h]
\begin{centering}
\includegraphics[width=10cm,height=6.5cm]{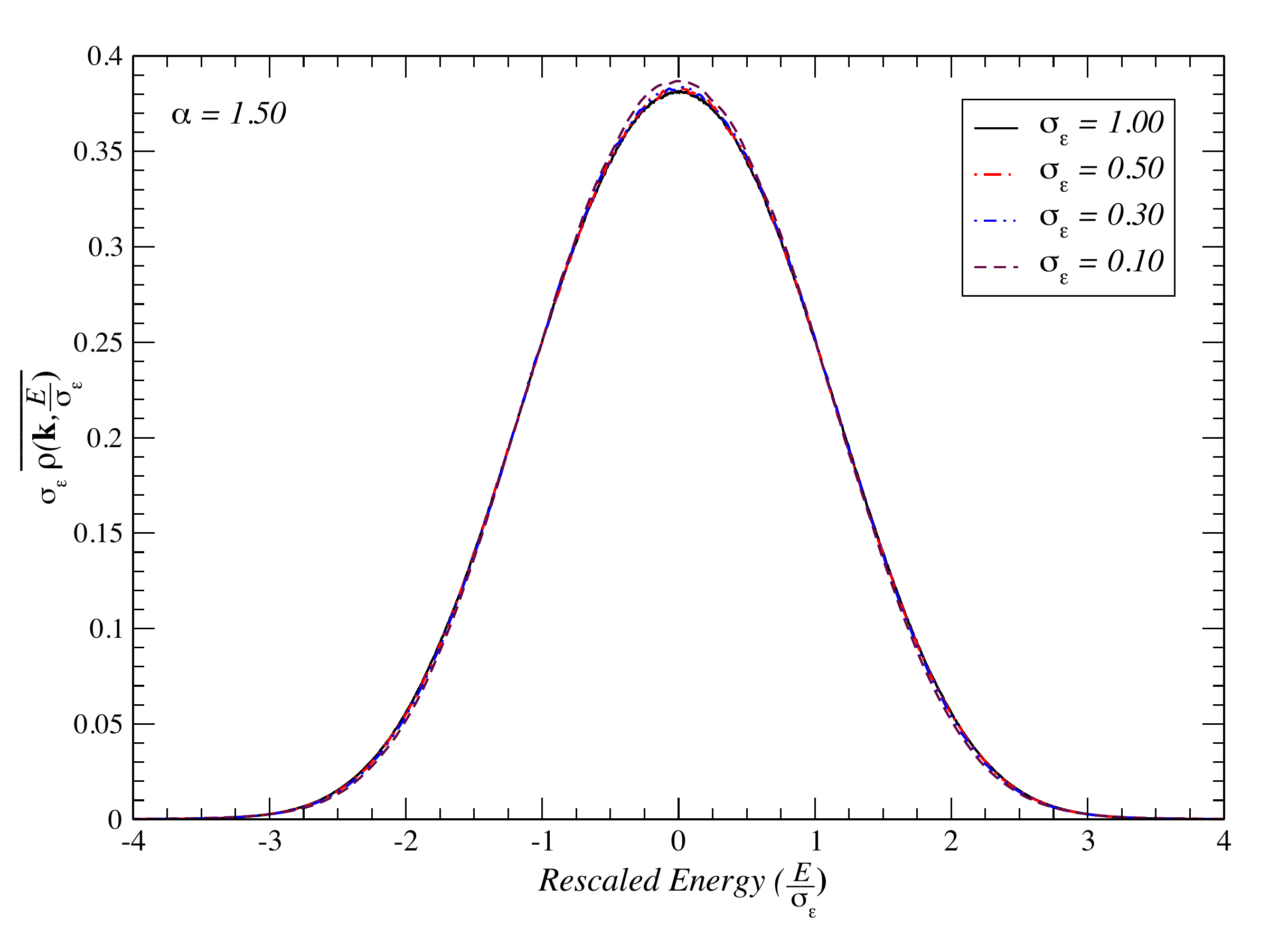} 
\par\end{centering}
\caption{The rescaled mean spectral function for the power-law correlated disorder,
in a system of size $L=2^{14}$ for different values of the on-site
variance $\sigma_{\varepsilon}$. There were used $8192$ Chebyshev
coefficients for the calculation. \label{fig:Scaled_PowerLaw} }
\end{figure}

Even though the form of the spectral function is not a gaussian, one
still observes (Figure~\ref{fig:Scaled_PowerLaw}) a universal behavior,
for different disorder strengths, similar to the one found for gaussian
disorder, namely 
\begin{equation}
\overline{\rho(k,E)}=\sigma_{\varepsilon}^{-1}\chi_{\alpha}\left(\frac{E}{\sigma_{\varepsilon}}\right)\label{eq:power_law_scale}
\end{equation}
with the $\chi_{\alpha}(\varepsilon$) depending on $\alpha$, but
not on the disorder variance $\sigma_{\varepsilon}$.

As for the gaussian disorder case, we will show that the results of
Figs\@.~\ref{fig:PowerLaw-SF}b and \ref{fig:Scaled_PowerLaw} reveal
the emergence of the classical limit, as a consequence of the local
character of scattering in momentum space.

\subsection{\label{subsec:Statistical-properties-of}Statistical properties of
the spectral function in the thermodynamic limit}

\begin{figure}[h]
\centering{}\includegraphics[width=13cm,height=7.1cm]{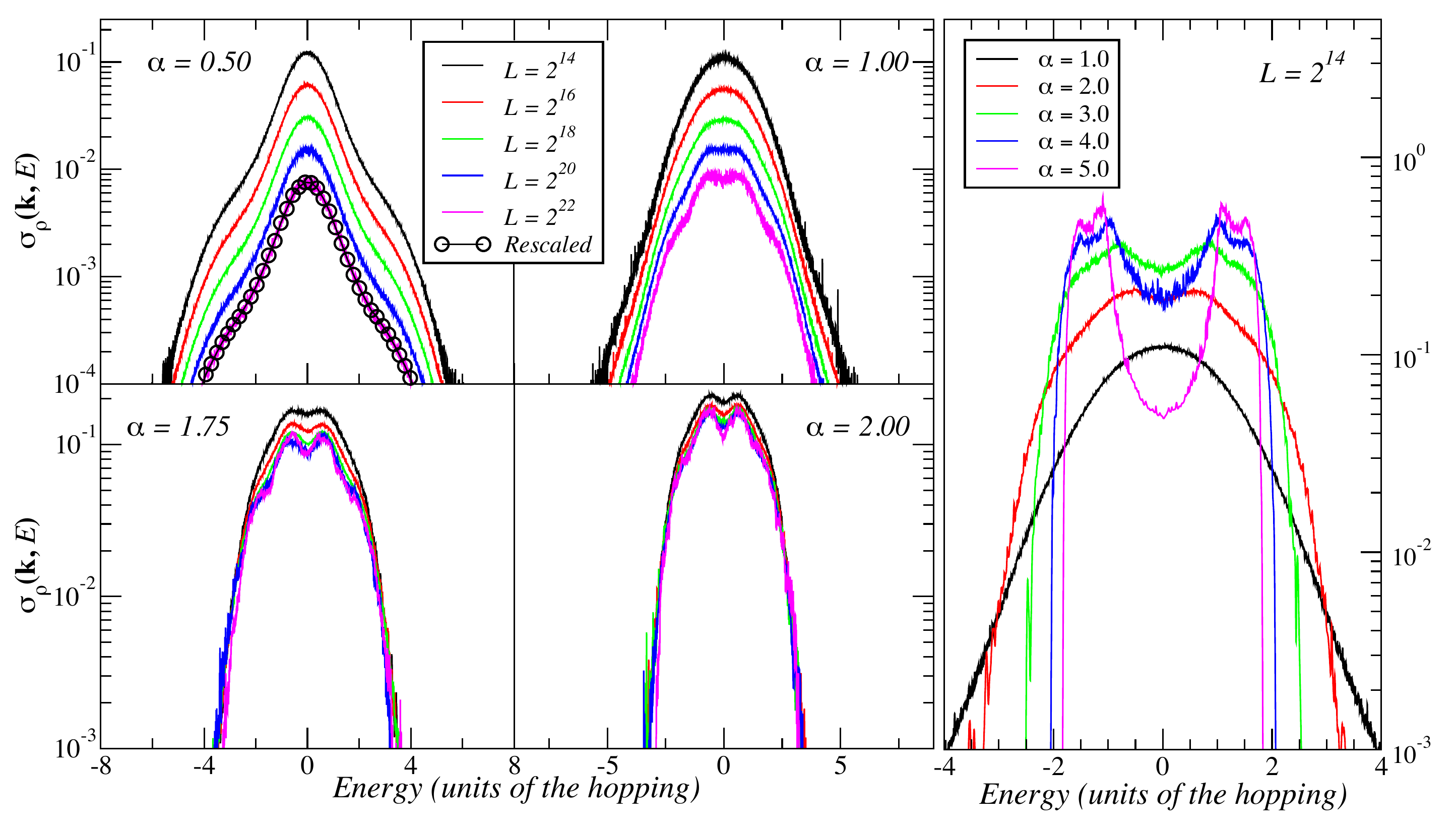}\caption{\label{fig:9}Plots of the standard deviation of the spectral function
at $k=\pm\frac{\pi}{2}$, for $\alpha=0.5$ (upper left panel), $\alpha=1.0$
(upper right panel), $\alpha=1.75$ (lower left panel) and $\alpha=2.0$
(lower right panel). For $\alpha=0.5$ the consecutive curves are
shown to collapse when rescaled by a factor of $L^{-\nicefrac{1}{2}}$
(black dots). For $\alpha>1$, the curves coalesce to a non-zero limiting
profile and no qualitative change of behavior is seen across $\alpha=2$.
In the extreme right panel, we show the decrease of the standard deviation
for larger values of $\alpha$ (at a fixed size). All the calculations
where done with $\sigma_{\varepsilon}=1$.}
\end{figure}

Thus far we have discussed the disorder-averaged spectral function.
It is not however clear if this quantity represents a typical value
for measurable quantity of macroscopic systems. This becomes specially
concerning in the case of the power-law disorder model, which is known
to have pathological properties in the thermodynamic limit\citep{ISI:000319729900004}.
To investigate this issue, we calculated the standard deviation of
$\rho(k,E)$ for increasing number of sites and different values of
the exponent $\alpha$. These results are shown for two examples in
Figure \ref{fig:9}.

From the numerical data, we conclude that for $\alpha<1$ the standard
deviation scales as $L^{-\nicefrac{1}{2}}$, which clearly indicates
a self-averaging behavior. On the other hand, for $\alpha>1$ there
seems to be a finite standard deviation for $\rho(k,E)$, even in
the thermodynamic limit, i.e. $\rho(k,E)$ still fluctuates from sample
to sample in the macroscopic limit. This property clearly indicates
that $\alpha=1$ is a special value for these models, not only because
the shape of $\overline{\rho(k,E)}$ changes, but also because it
becomes a non-self-averaging quantity.

\begin{figure}[h]
\centering{}\includegraphics[width=10cm,height=6.5cm]{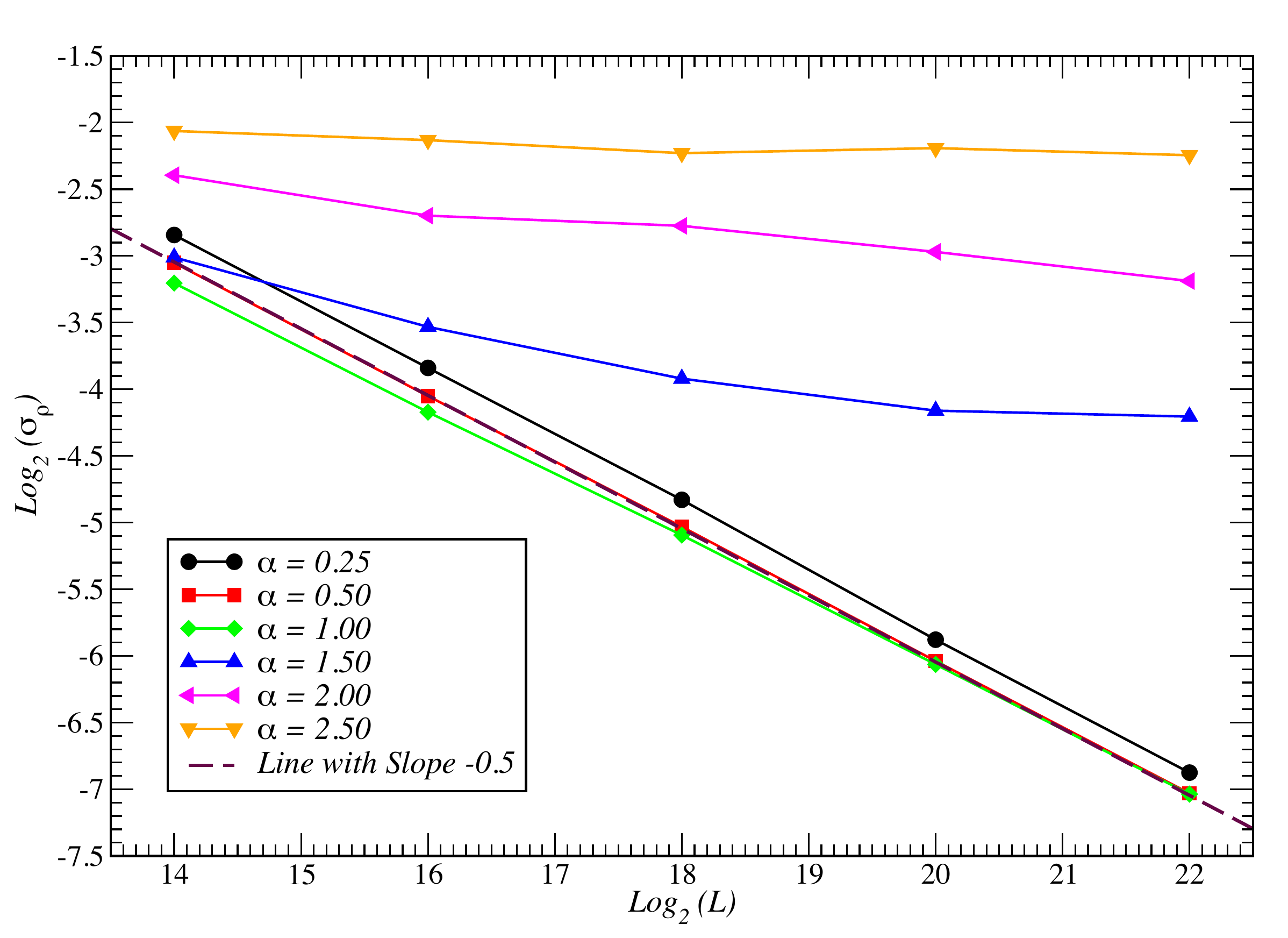}\caption{\label{fig:Scaling-of-the}Scaling of the standard deviation of $\rho(k,E)$
for $k=\frac{\pi}{2}$ and $E=0$, as a function of the system's size.
The dashed line stands for the usual $L^{-\nicefrac{1}{2}}$ scaling.}
\end{figure}

In Figure \ref{fig:9}, we can also see an example of the same calculation
done for $\alpha=2$, where no qualitative changes in the scaling
behavior of $\sigma_{\rho}$ can be seen. Obviously, for very large
values of $\alpha$, these persistent fluctuations start to decrease,
since the system is approaching an ordered limit ($\alpha\rightarrow+\infty$).
To sum up these results, we present in Figure \ref{fig:Scaling-of-the},
a plot showing the scaling of $\sigma_{\rho}$ at the central energy,
with the increase in the chain size.

\section{Analytical Results and Discussion\label{sec:Analysis-and-discussion}}

If the state at $t=0$ is $\ket{\psi(0)}=\ket k$, the probability
amplitude that the state at time $t$ is still the same is $\bra ke^{-i\mathcal{H}t/\hbar}\ket k$.
Using a complete set of energy eigenstates $\left\{ \ket{\psi_{\beta}}:\text{\ensuremath{\beta=0,\dots,N-1}}\right\} $,
we can see that this amplitude is the Fourier transform of the spectral
function defined in Eq~.\ref{eq:SDOS}: 
\begin{align}
\bra ke^{-i\mathcal{H}t/\hbar}\ket k & =\sum_{\beta}e^{-iE_{\beta}t/\hbar}\left|\braket{\psi_{\beta}}k\right|^{2}\nonumber \\
 & =\int_{-\infty}^{+\infty}dEe^{-iEt/\hbar}\sum_{\beta}\left|\braket{\psi_{\beta}}k\right|^{2}\delta\left(E-E_{\beta}\right)\nonumber \\
 & =\int_{-\infty}^{+\infty}dEe^{-iEt/\hbar}\rho(k,E).\label{eq:FT_spectral}
\end{align}
Expanding both sides in powers of $t$ and averaging over disorder,
we get the following expression for the $n^{th}$-moment of the disorder-averaged
spectral function $\overline{\rho(k,E)}$: 
\begin{equation}
\overline{\bra k\mathcal{H}^{n}\ket k}=\int_{-\infty}^{+\infty}dEE^{n}\overline{\rho\left(k,E\right)}.\label{eq:moments}
\end{equation}
The Hamiltonian is the one defined in Eq.~\ref{eq:Hamiltonian} and
can be written as $\mathcal{H}=\mathcal{H}_{0}+\mathcal{V}$ where
\begin{subequations} 
\begin{align}
\mathcal{H}_{0}= & \sum_{k}E_{k}\ket k\bra k\\
\mathcal{V}= & \sum_{m}\varepsilon_{m}\ket{\varphi_{m}}\bra{\varphi_{m}},
\end{align}
\end{subequations} with the band Hamiltonian $\mathcal{H}_{0}$ being
diagonal in the Bloch basis, and the disordered potential, $\mathcal{V},$
in the local Wannier basis. In the calculation of $\bra k\mathcal{H}^{n}\ket k$,
we will assume that $\mathcal{H}_{0}\ket k=E_{k}\ket k=0$. This is
strictly true for the states in the center of the band (i.e $k=\pm\pi/2$),
for which we calculated numerically the spectral function. However,
this assumption implies no loss of generality, since for an arbitrary
value $k$, we can add an irrelevant constant to $\mathcal{H}$, 
\begin{equation}
\mathcal{H}_{0}\to\mathcal{H}_{0}:=\sum_{k'}\left(E_{k'}-E_{k}\right)\ket{k'}\bra{k'},
\end{equation}
such that $\mathcal{H}_{0}\ket k=0$, remains true. The calculation
will show that changing $k$ only shifts the spectral function in
energy, by the value of $E_{k}$.

\subsection{Gaussian case}

As a justification for our numerical results, we managed to calculate
the average spectral function for the infinite chain, with a gaussian
model of correlated disorder. Generally, our analytical results will
be valid in the limits when $2\pi/L\ll q_{c}\ll\pi$ and $q_{c}\ll\sigma_{\varepsilon}/\hbar v_{k}$.

\subsubsection{Lowest Order Terms}

To illustrate the gist of the argument, we begin by looking at the
lowest order moments, using the Eq.~\ref{eq:moments}.

It is obvious that for $n=1$ the result is zero, because $\mathcal{H}_{0}\ket k=0$
and $\overline{\mathcal{V}}=0$. For $n=2,$ 
\begin{align}
\overline{\bra k\mathcal{H}^{2}\ket k}= & \overline{\bra k(\mathcal{H}_{0}+\mathcal{V})(\mathcal{H}_{0}+\mathcal{V})\ket k}\nonumber \\
= & \overline{\bra k\mathcal{V}^{2}\ket k},
\end{align}
and resolving the identity in the Bloch basis, 
\begin{equation}
\overline{\bra k\mathcal{H}^{2}\ket k}=\sum_{q}\overline{\bra k\mathcal{V}\ket{k+q}\bra{k+q}\mathcal{V}\ket k}.
\end{equation}
Recalling Eq. \ref{eq:disorder_corr}, 
\begin{equation}
\overline{\bra k\mathcal{H}^{2}\ket k}=\sum_{q}V^{2}(q)=\sigma_{\varepsilon}^{2},
\end{equation}
By the same arguments, in the third moment only one term survives:
\begin{align}
\overline{\bra k\mathcal{H}^{3}\ket k} & =\overline{\bra k\mathcal{V}\mathcal{H}_{0}\mathcal{V}\ket k}\nonumber \\
 & =\sum_{q}V^{2}(q)E_{k+q};\label{eq:3rd order}
\end{align}
In the thermodynamic limit, the sum over $q$ turns into an integral
and if $q_{c}\ll\pi$, we can extend the integration range to $q\in\left]-\infty,\infty\right[$
and expand $E_{k+q}\approx\hbar v_{k}q$. In this case, the integrand
is odd in $q$ and the right-hand side of Eq.$\;$\ref{eq:3rd order}
vanishes upon integration.

Finally, we tackle the $4^{th}$-moment (the last, before presenting
the general argument), whose the only non-zero terms are 
\begin{equation}
\overline{\bra k\mathcal{\mathcal{H}}^{4}\ket k}=\overline{\bra k\mathcal{\mathcal{V}}\mathcal{H}_{0}^{2}\mathcal{V}\ket k}+\overline{\bra k\mathcal{\mathcal{V}}^{4}\ket k}.
\end{equation}
Using the same technique as above, the first term is 
\begin{equation}
\sum_{q}V^{2}(q)E_{k+q}^{2}=\sum_{q}V^{2}(q)\left(\hbar v_{k}q\right)^{2},\label{eq:h_0_contribs}
\end{equation}
which is a complete gaussian integral (in the limit $q_{c}\ll\pi$),
whose value is 
\begin{equation}
\bra k\mathcal{\mathcal{V}}\mathcal{H}_{0}^{2}\mathcal{V}\ket k=\sigma_{\varepsilon}^{2}\left(\hbar v_{k}q_{c}\right)^{2}.\label{eq:55}
\end{equation}
On the other hand, the term containing the $4^{th}$ power of $\mathcal{V}$
is 
\begin{align*}
\overline{\bra k\mathcal{\mathcal{V}}^{4}\ket k} & =\sum_{q_{1},q_{2},q_{3}}V(q_{1})V(q_{2})V(q_{3})V(-q_{1}-q_{2}-q_{3})\\
 & \times\overline{e^{i\phi_{q_{1}}}e^{i\phi_{q_{2}}}e^{i\phi_{q_{3}}}e^{i\phi_{-q_{1}-q_{2}-q_{3}}}},
\end{align*}
The averages of these random phase factors are discussed in the \ref{sec:Averages}.
In particular, we show that, in the thermodynamic limit ($L\rightarrow\inf$),
the expression above reduces to 
\begin{equation}
\overline{\bra k\mathcal{\mathcal{V}}^{4}\ket k}=3\left[\sum_{q}V^{2}\left(q\right)\right]^{2}=3\sigma_{\varepsilon}^{4}.\label{eq:56}
\end{equation}
Finally, by looking at the Eqs.$\;$\ref{eq:55} and \ref{eq:56},
we see that, as long as $\sigma_{\varepsilon}^{2}\gg\left(\hbar v_{k}q_{c}\right)^{2}$,
we can ignore terms that have insertions of $\mathcal{H}_{0}$. Then,
we simply write $\overline{\bra k\mathcal{\mathcal{H}}^{4}\ket k}$
as: 
\begin{equation}
\overline{\bra k\mathcal{\mathcal{H}}^{4}\ket k}\approx\overline{\bra k\mathcal{\mathcal{V}}^{4}\ket k}=3\sigma_{\varepsilon}^{4}
\end{equation}

\subsubsection{General Expression for the Moments of $\rho(k,E)$}

Inspired on the results above, we argue that the general form of the
terms in Eq.$\:$\ref{eq:moments} is:

\begin{equation}
\overline{\bra k\mathcal{\mathcal{H}}^{2p}\ket k}\approx\overline{\bra k\mathcal{\mathcal{V}}^{2p}\ket k},\label{eq:classical_result}
\end{equation}

\begin{equation}
\overline{\bra k\mathcal{\mathcal{H}}^{2p+1}\ket k}\approx0.\label{eq:classical_result-1}
\end{equation}

Furthermore, in the \ref{sec:Averages} we show that the averages
$\overline{\bra k\mathcal{V}^{2p}\ket k}$ have the following general
form

\begin{equation}
\overline{\bra k\mathcal{V}^{2p}\ket k}=(2p-1)!!\left(\sigma_{\varepsilon}^{2}\right)^{p}\left[1+\mathcal{O}\left(\frac{1}{L}\right)\right].\label{eq:60}
\end{equation}
Using the Eqs.$\;$\ref{eq:classical_result}-\ref{eq:60}, in the
thermodynamic limit ($L\rightarrow\inf$), we can rebuild the entire
Taylor series for the averaged diagonal propagator, and re-sum it
as follows: 
\begin{align*}
\overline{\bra ke^{-i\mathcal{H}t/\hbar}\ket k} & =\sum_{p=0}^{\infty}\frac{1}{(2p)!}\left(\frac{-it}{\hbar}\right)^{2p}\overline{\bra k\mathcal{\mathcal{V}}^{2p}\ket k}\\
 & =\sum_{p=0}^{\infty}\frac{(-1)^{p}(2p-1)!!}{(2p)!}\left(\frac{\sigma_{\varepsilon}^{2}t^{2}}{\hbar^{2}}\right)^{p}\\
 & =\sum_{p=0}^{\infty}\frac{1}{2^{p}p!}\left(-\frac{\sigma_{\varepsilon}^{2}t^{2}}{\hbar^{2}}\right)^{p}=e^{-\sigma_{\varepsilon}^{2}t^{2}/2\hbar^{2}}
\end{align*}
The spectral function is the time-domain Fourier transform of this
last expression, yielding 
\begin{equation}
\overline{\rho(k=\pm\frac{\pi}{2},E)}=\frac{1}{\sqrt{2\pi\sigma_{\varepsilon}^{2}}}e^{-\frac{E^{2}}{2\sigma_{\varepsilon}^{2}}},\label{eq:gaussian-1}
\end{equation}
which agrees with the results found in our numerical calculations,
using the KPM.

For the sake of completeness, we also state the result for a general
value of $k$, which can be obtained from Eq.$\:$\ref{eq:gaussian-1}
simply by shifting the energy variable by the corresponding band energy
$E_{k}$ of that state, i.e.

\begin{equation}
\overline{\rho(k,E)}=\frac{1}{\sqrt{2\pi\sigma_{\varepsilon}^{2}}}e^{-\frac{\left(E-E_{k}\right)^{2}}{2\sigma_{\varepsilon}^{2}}}.\label{eq:gaussian-1-1}
\end{equation}

In conclusion, we found that, if $q_{c}\ll\pi$ and $\left(v_{k}q_{c}\right)^{2}\ll\sigma_{\varepsilon}^{2}$,
then the disorder-averaged spectral function, in the thermodynamic
limit, will have a gaussian shape. This is true, even if the disorder
strength (measured by $\sigma_{\varepsilon}$) is small, as long as
this is matched by a decrease of $q_{c}$ and corresponding increase
of the correlation length of the potential. For instance, the mean
free path, estimated by $\ell=\hbar v_{k}/\sigma_{\varepsilon}$ can
still be much larger that the lattice parameter, so long as $\ell<\xi$
, where $\xi$ is the disorder correlation length.

\subsubsection{Emergence of the Classical Limit for the Spectral Function}

We were able to establish precise conditions in which the classical
limit of the spectral function, found by Trappe \emph{et.} \emph{al}.\citep{ISI:000355253100005},
appears. The statement of this limit is equivalent to Eq.~\ref{eq:classical_result},
and reads ($E_{k}=0$) 
\begin{equation}
\overline{\bra ke^{-i\mathcal{H}t/\hbar}\ket k}=\overline{\bra ke^{-i\mathcal{V}t/\hbar}\ket k}.
\end{equation}
so that 
\begin{equation}
\overline{\rho(k,E)}=\int dte^{iEt/\hbar}\overline{\bra ke^{-i\mathcal{V}t/\hbar}\ket k}.
\end{equation}
Using the Wannier basis (eigenbasis of $\mathcal{V}$) and its transformation
law to the Bloch basis $\braket{\varphi_{n}}k=\exp\left(ikn\right)/\sqrt{L}$,
we can rewrite the above equation (with $E_{k}=0$) as

\begin{align}
\overline{\bra ke^{-i\mathcal{V}t/\hbar}\ket k} & =\sum_{n,m}\overline{\braket k{\varphi_{n}}\bra{\varphi_{n}}e^{-i\mathcal{V}t/\hbar}\ket{\varphi_{m}}\braket{\varphi_{m}}k}\nonumber \\
 & =\frac{1}{L}\sum_{m}\overline{e^{-i\varepsilon_{m}t/\hbar}}=\int dEP(E)e^{-iEt/\hbar},
\end{align}
where $P(E)$ is the probability distribution of a site energy. Comparing
the above with Eq.~\ref{eq:FT_spectral}, we have

\begin{equation}
\overline{\rho(k,E)}=P(E).\label{eq:classical_spectralFuntion}
\end{equation}
Thus, the averaged spectral function is just the probability distribution
of a single site energy. As it is clear for the definition of the
disordered potential (Eq.$\;$\ref{eq:corr pot}), the distribution
$P(E)$ must be a gaussian according of the Central Limit Theorem.

\subsection{Power-Law Correlated Disorder}

\subsubsection{Validity of the Classical Limit}

In the case of Power-law correlated disorder, the argument leading
to the Eq.~\ref{eq:classical_result} still holds, as long as $\alpha>1$,
but requires a slightly different formulation. To see how this comes
about, let us consider Eq.~\ref{eq:h_0_contribs} as an example.
In this case, we have 
\begin{align}
\sum_{q}V^{2}(q)E_{k+q}^{2} & =\frac{2\pi}{L}A^{2}(\alpha)\sum_{q\ne0}\frac{1}{\left|q\right|^{\alpha}}\left[E_{k+q}\right]^{2}\nonumber \\
 & =\frac{\sigma_{\varepsilon}^{2}}{2\mathcal{\zeta}(\alpha)}\sum_{p=1}^{L/2}\frac{1}{p^{\alpha}}E_{k+2\pi p/L}^{2}.
\end{align}
As before, if we expand $E_{k+q}$ in powers of $p$, we get terms
of the form 
\begin{equation}
\left[\frac{1}{n!}\frac{d^{n}E_{k}}{dk^{n}}\right]\frac{\sigma_{\varepsilon}^{2}}{2\mathcal{\zeta}(\alpha)}\left(\frac{2\pi}{L}\right)^{n}\sum_{p=1}^{L/2}\frac{1}{p^{\alpha-n}}
\end{equation}
If $\alpha-n>1$ the sum above is convergent and the result vanishes,
in the large-$L$ limit, as $L^{-n}.$ On the other hand, if $\alpha-n<1$
the sum diverges, but instead it can be written as an integral over
the First Brillouin Zone, as follows 
\begin{align}
\left(\frac{2\pi}{L}\right)^{n}\sum_{p=1}^{L/2}\frac{1}{p^{\alpha-n}} & =\left(\frac{2\pi}{L}\right)^{\alpha-1}\int_{\frac{2\pi}{L}}^{\pi}dq\frac{1}{q^{\alpha-n}}\nonumber \\
= & \frac{1}{n-\alpha+1}\left[\frac{2^{\alpha-1}\pi^{n}}{L^{\alpha-1}}-\left(\frac{2\pi}{L}\right)^{n}\right].
\end{align}
Both terms in the equation above go to zero in the thermodynamic limit,
since $\alpha>1$ and $n\geq1$. This argument is obviously true for
every term in $\bra k\mathcal{H}^{n}\ket k$, containing insertions
of $\mathcal{H}_{0}$. Hence, in the $L\rightarrow\infty$ limit,
the only finite contributions come from the all-$\mathcal{V}$ terms,
and we re-obtain the classical result expressed in Eq.~\ref{eq:classical_spectralFuntion}.

In this limit the spectral function can only depend on the parameters
of the disordered potential, namely $\sigma_{\varepsilon}$ and $\alpha$.
Since $\alpha$ is dimensionless, there is a single energy scale,
$\sigma_{\varepsilon}$, in $\overline{\rho(k,E)}$. The scaling of
Eq.~\ref{eq:power_law_scale}, illustrated in Figure~\ref{fig:Scaled_PowerLaw},
follows at once. It should be noted, however, that as $\alpha$ gets
closer to 1, this scaling is not observed numerically. This is due
to finite size effects that we have not accounted for. An example
is the very slow convergence of $\sum_{p=1}^{L/2}p^{-\alpha}$ to
$\zeta(\alpha)$. For $\alpha=1.1,$ for instance, the truncation
error is still of order 10\% for $L\sim10^{10}.$

\subsubsection{The Limiting Cases ($\alpha\rightarrow1$ and $\alpha\rightarrow+\protect\inf$)
And The Double-Peaked Shape }

Despite the validity of the classical limit for the averaged spectral
function, we have shown in the \ref{sec:Averages} that it is not
clear how to obtain a closed form for the $n^{th}$-moment of $\overline{\rho(k,E)}$
even in this limit. Nevertheless, the limit $\alpha\rightarrow1^{+}$
revealed itself as very special case, where the exact averaged spectral
function is found to be a gaussian,

\begin{equation}
\overline{\rho(k,E)}=\frac{1}{\sqrt{2\pi\sigma_{\varepsilon}^{2}}}e^{-\frac{\left(E-E_{k}\right)^{2}}{2\sigma_{\varepsilon}^{2}}}.\label{eq:gaussian-1-1-1}
\end{equation}
This result is consistent with the numerical results obtained in the
last section (see Figure$\;$\ref{fig:PowerLaw-SF}).

For $\alpha>1$, however, the higher cumulants of the spectral function
cease to be zero, and $\overline{\rho(k,E)}$ drifts away from a gaussian
shape. For illustration, we have calculated the $4^{th}$-cumulant
of the averaged spectral function, as a function of the exponent $\alpha$
. This has the following definition:

\begin{equation}
m_{4}=\int_{-\infty}^{+\infty}dEE^{4}\overline{\rho\left(k,E\right)}-3\left[\int_{-\infty}^{+\infty}dEE^{2}\overline{\rho\left(k,E\right)}\right]^{2},
\end{equation}

and can be directly computed using the expressions obtained in the
\ref{sec:Averages}, \emph{i.e}.

\begin{equation}
m_{4}(\alpha)=-3\sigma_{\varepsilon}^{4}\frac{\zeta(2\alpha)}{2\zeta(\alpha)}.\label{eq:4thcumul}
\end{equation}

Other than explaining the deviations from the gaussian shape that
we found in the numerical plots of $\overline{\rho(k,E)}$, these
effects have another striking consequence. According to our earlier
remarks, in the classical limit, the averaged spectral function is
the same as the probability distribution of the site energies. Since
the value of the disordered potential in a single point is described
as a sum of a large number of independent random variables (see Eq.$\;$\ref{eq:power-law-disorder}),
the non gaussian shape shows that these do not obey the \textbf{Central
Limit Theorem}. To see how this comes about, we start by looking at
Eq.$\;$\ref{eq:24}, where 
\begin{equation}
\sigma_{\varepsilon}^{2}\propto\sum_{p=1}^{L/2}\frac{1}{p^{\alpha}}.
\end{equation}

When $\alpha>1$, this sum is convergent in the $L\to\infty$ limit,
which means that only a number of $\mathcal{O}(1)$ of terms actually
contribute to the variance of the local disorder $\varepsilon_{n}$.
Furthermore, as $\alpha$ increases, this sum is dominated by less
and less terms, meaning that we are never in the conditions of the
central limit theorem (which assumes a large number of summed random
independent variables). 
\begin{figure}[h]
\begin{centering}
\includegraphics[width=10cm,height=6.9cm]{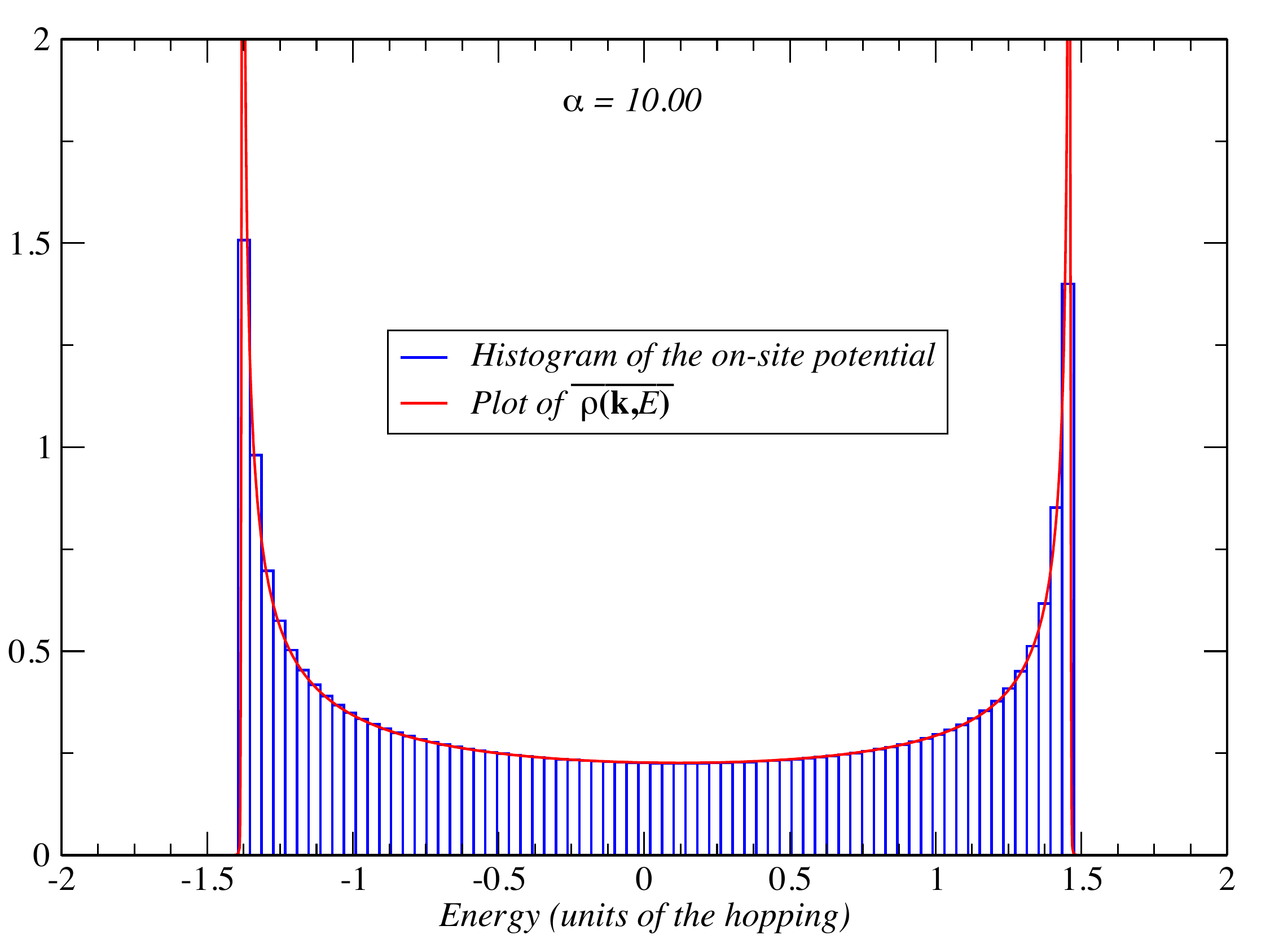} 
\par\end{centering}
\caption{Comparison of the local energy distribution $P(E)$ and the disorder-averaged
spectral function $\overline{\rho(k,E)}$, obtained for a system of
size $L=2^{17}$ with $8192$ Chebyshev expansion coefficients and
a single realization of disorder. The calculation was done for $k=\nicefrac{\pi}{2}$.\label{fig:pdistribution} }
\end{figure}

\begin{figure}[h]
\begin{centering}
\includegraphics[width=10cm,height=7.2cm]{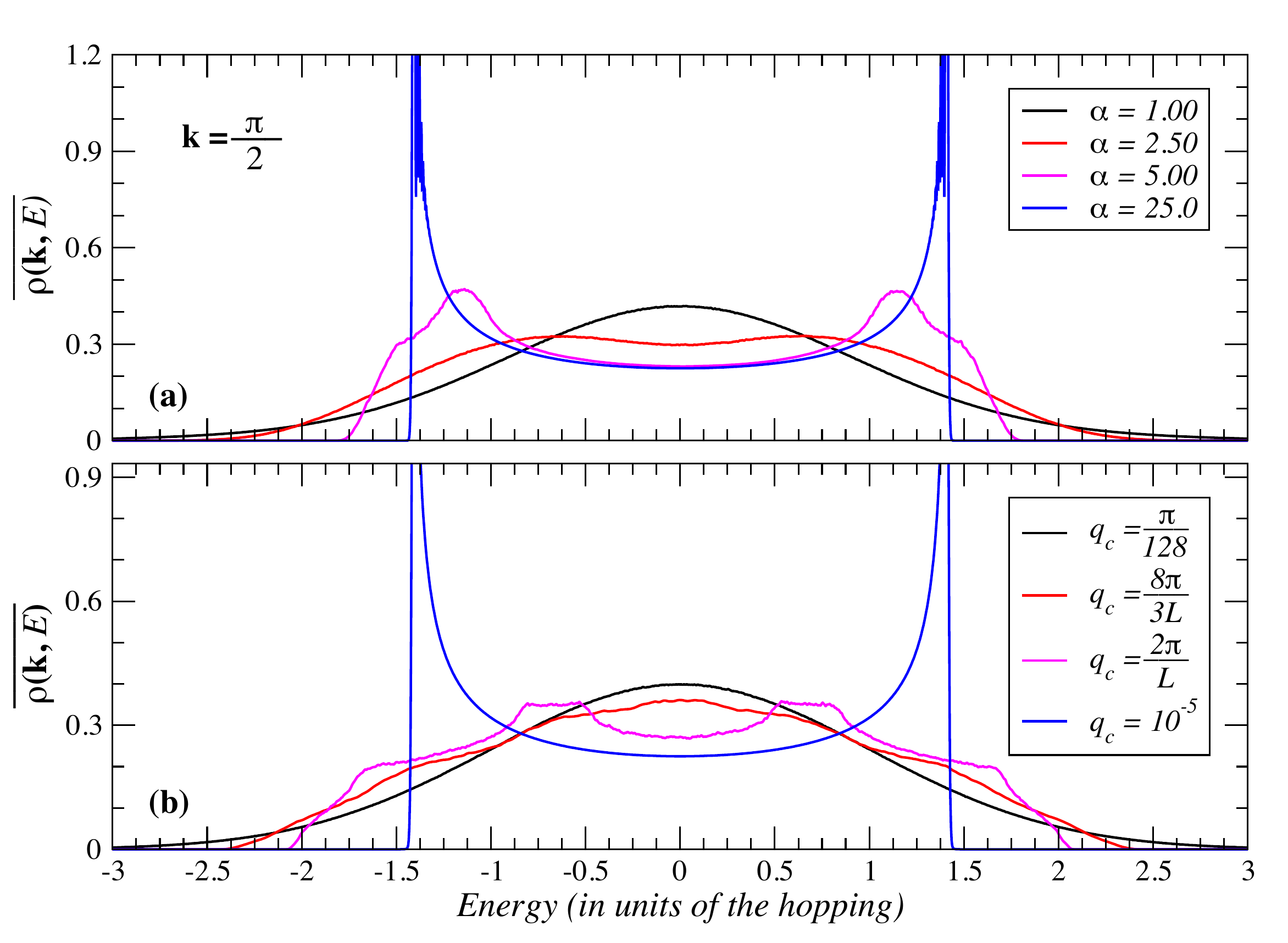} 
\par\end{centering}
\caption{Spectral function $\rho(\mathbf{k},E)$ for the disordered system
of size $L=2^{14}$ with $8192$ Chebyshev moments for different values
of $\alpha$ (top panel) and $q_{c}$ (lower panel). The limits $\alpha\gg1$
and $q_{c}\ll2\pi/L$ are identical: see text . \label{fig:Alpha_large_Spectral-Function}}
\end{figure}

This becomes particularly clear in the extreme case $\alpha\rightarrow+\inf$.
In this limit, the local value of the disordered potential is dominated
by a single term, $p=1$, and the disorder is a static cosine potential
with a wavelength $L$ and a random phase, 
\begin{equation}
\varepsilon_{n}\sim\sqrt{2}\sigma_{\varepsilon}^{2}\cos(\frac{2\pi n}{L}+\phi_{2\pi/L}).
\end{equation}
The corresponding probability density function can be calculated,
yielding the expression:

\begin{equation}
P(E)=\frac{1}{\pi}\frac{1}{\sqrt{2\sigma_{\varepsilon}^{2}-E^{2}}}.\label{eq:69}
\end{equation}
As an illustration, we depict in Figure~\ref{fig:pdistribution}
the KPM calculated the spectral function for $\alpha=10,$ and the
normalized histogram of site energies for a single realization of
disorder. As $\alpha$ increases above $1$, the spectral function
smoothly approaches the limiting form of Eq.~\ref{eq:69}, by first
displaying a two peaked shape as illustrated in Figure~\ref{fig:Alpha_large_Spectral-Function}a.

The expression of Eq.~\ref{eq:69} also corresponds to the one we
obtain numerically for gaussian disorder case when $q_{c}\ll2\pi/L$
(see Figure~\ref{fig:Alpha_large_Spectral-Function}b). In either
case, of course, a single value $q$ dominates the sum 
\begin{equation}
\varepsilon_{n}=2\sum_{q>0}V(q)\cos\left(qn+\phi_{q}\right)
\end{equation}
and the two models of disorder cannot be distinguished. 

\section{Conclusions}

We have studied the spectral function of Bloch states in a tight-binding
chain, with two models of correlated disorder: the gaussian model
(with a correlation length given by $q_{c}^{-1}$) and the power-law
model (with an algebraic decay of correlations characterized by an
exponent $\alpha$). For both models, we calculated numerically (with
KPM), and analytically, in certain limits, the disorder-averaged single-particle
spectral function $\overline{\rho(k,E)}$, at zero temperature. We
also evaluated numerically the fluctuations of this quantity from
sample to sample, in the power-law model, in order to study its self-averaging
character.

The analytical calculations of $\overline{\rho(k,E)}$ were done in
the thermodynamic limit, by resuming the short-time expansion of the
diagonal propagator in momentum space. For the gaussian case, we found
out that, in the regimes when $q_{c}\ll\sigma_{\varepsilon}/\hbar v_{k}$,
and the correlation length of the disorder is much larger than the
lattice spacing ($q_{c}a\ll1)$ but much smaller than the system's
size, the spectral function has a gaussian shape, $\rho(k,E)$ with
mean $\mu=E_{k}$ and variance $\sigma_{\varepsilon}^{2}$, the variance
of the random site energy. This is consistent with the classical limit
for the propagator\citep{ISI:000355253100005} applied to our lattice
system.

In the power-law model, where there is no energy scale associated
with the space-correlations, we still found that the averaged spectral
function is given by its classical limit, but only if the exponent
$\alpha$, characterizing the algebraic decay of the power-spectrum,
exceeds unity (while the delocalization of the eigenstates\citep{ISI:000076616000046}
occurs only at $\alpha=2$). The mean spectral function is a gaussian
in the limit $\alpha\rightarrow1^{+}$, but develops non-zero higher
cumulants for larger values of $\alpha$, reflecting the actual distribution
of on-site energies. The spectral density follows a scaling law similar
to the one found for the gaussian disorder case. Although we are unable
to find an exact functional form for $\overline{\rho(k,E)}$, this
scaling law can be understood from the fact that there are no other
energy scales in the problem besides $\sigma_{\varepsilon}$ (since
$\alpha$ is a dimensionless parameter); hence, $\sigma_{\varepsilon}\overline{\rho(k,E)}$
must be a function of $E/\sigma_{\varepsilon}$. All these results
are confirmed by our numerical calculations of $\overline{\rho(k,E)}$.

For the later model, we discovered that the standard deviation of
the spectral function, for $\alpha\geq1$, does not go to zero in
the thermodynamic limit. This means that in the non-perturbative regime,
the spectral function is not a self-averaging quantity and remains
sample dependent in the infinite size system. While this may not come
as a surprise in such a pathological model, it also reinforces that
$\alpha=1$ is a crossover point for these potentials. More surprisingly,
the results on the single-particle spectral function do not seem to
give any indication that $\alpha=2$ is a special point for these
models, as was argued by Petersen et al\citep{ISI:000319729900004}
in relation to the predicted delocalization transition. Granted that
there is no obvious relation between the spectral function and the
localization/delocalization of the eigenstates, one could still expect
that a qualitative change in the disordered potential might show up
at the transition point. Yet, we found no such effect.

\,

In conclusion, we studied the spectral function in a 1D band model
with correlated disorder. Through a combination of numerical and analytical
work we were able to obtain results in a non-perturbative regime,
and show explicitly how the classical limit of the spectral function
emerges \citep{ISI:000355253100005}. In the case of power-law disorder,
this happens when the local distribution of site energies is not gaussian,
due to inapplicability of the central limit theorem. The localization
transition in these models occurs deep in the region where the spectral
function is classical, and that raises the question of whether something
may be learned on that transition from this knowledge of the spectral
function.

\section{Acknowledgments}

For this work, N. A. Khan was supported by the grants ERASMUS MUNDUS
Action 2 Strand 1 Lot 11, EACEA/42/11 Grant Agreement 2013-2538 /
001-001 EM Action 2 Partnership Asia-Europe and research scholarship
UID/FIS/04650/2013 of Fundação da Ciência e Tecnologia; J. P. Santos
Pires was supported by the MAP-fis PhD grant PD/BD/142774/2018 of
Fundação da Ciência e Tecnologia.

The work at Centro de Física do Porto, as a whole, is supported by
the grant UID/FIS/04650/2013 of Fundação da Ciência e Tecnologia.

We would also like to thank the referee for its suggestions about
the pathological properties of the power-law correlated model, which
drove us to extend the paper with the data shown in the Subsection
\ref{subsec:Statistical-properties-of}.

\appendix
%dummy comment inserted by tex2lyx to ensure that this paragraph is not empty%dummy comment inserted by tex2lyx to ensure that this paragraph is not empty

\section{Random Phase Averages\label{sec:Averages}}

In section \ref{sec:Analysis-and-discussion}, we needed to calculate
terms of the form

\begin{equation}
\overline{\exp\left(i\phi_{q_{1}}\right)\exp\left(i\phi_{q_{2}}\right)\dots\exp\left(i\phi_{-q_{1}-q_{2}-\dots q_{n-1}}\right)}
\end{equation}

where $\phi_{q}$ are independent random phases with an uniform distribution
in the circle and obeying the constraint $\phi_{q}=-\phi_{q}$. These
expressions appear inside sums over momenta, of the form 
\begin{align}
\sum_{q_{1}\neq0}\dots\sum_{q_{n-1}\neq0}V(q_{1})\dots V(q_{n-1})V(-q_{1}\dots-q_{n-1})\nonumber \\
\times\overline{\exp\left(i\phi_{q_{1}}\right)\dots\exp\left(i\phi_{-q_{1}-q_{2}-\dots q_{n-1}}\right)} & ,\label{eq:V_sums}
\end{align}
where $V(q)=V(-q)$.

Clearly, since these phases are uniformly distributed independent
variables (except in the case $q_{2}=\pm q_{1}$), we have\begin{subequations}
\begin{align}
\overline{\exp\left(i\phi_{q_{1}}\right)} & =0,\label{eq:average1}\\
\overline{\exp\left(i\phi_{q_{1}}\right)\exp\left(i\phi_{q_{2}}\right)} & =\delta_{q_{1}+q_{2},0}.\label{eq:average_pair}
\end{align}
\end{subequations}Therefore, we can only obtain a non zero result
if all the phase factors are paired. This means that $F(q_{1},\dots,q_{n})=\overline{\exp\left(i\phi_{q_{1}}\right)\exp\left(i\phi_{q_{2}}\right)\dots\exp\left(i\phi_{q_{n}}\right)}$
is zero unless $\sum_{i}q_{i}=0$.

\subsection*{General Procedure}

To actually calculate the phase averages, we may start with the following
illustrative case: 
\begin{equation}
F\left(q_{1},q_{2},q_{3},q_{4}\right):=\overline{\exp\left(i\phi_{q_{1}}\right)\exp\left(i\phi_{q_{2}}\right)\exp\left(i\phi_{q_{3}}\right)\exp\left(i\phi_{q_{4}}\right)}
\end{equation}
To prevent lengthy notation, we define 
\begin{align*}
\delta_{q_{i}+q_{j},0} & \to\delta_{ij}\\
1-\delta_{q_{i}+q_{j},0} & \to\overline{\delta_{ij}}=1-\delta_{ij}
\end{align*}
such that $\delta_{ij}+\overline{\delta_{ij}}=1$. Note also, that
since $V(q)=V(-q)$, the contraction of two momenta is equivalent
to a Kronecker delta in the momentum sums.

Hence, we can write 
\begin{align*}
F(q_{1},\dots,q_{4}) & =\delta_{12}F(q_{3},q_{4})+\overline{\delta_{12}}F(q_{1},\dots,q_{4})\\
 & =\delta_{12}\delta_{34}+\overline{\delta_{12}}F(q_{1},\dots,q_{4}),
\end{align*}
and repeat the process until we exhaust all possibilities. In this
case, we just need to do it once, 
\begin{align*}
\overline{\delta_{12}}F(q_{1},\dots,q_{4}) & =\overline{\delta_{12}}\left[\delta_{13}\delta_{24}+\overline{\delta_{13}}F(q_{1},\dots,q_{4})\right]\\
 & =\overline{\delta_{12}}\left[\delta_{13}\delta_{24}+\overline{\delta_{13}}\delta_{14}\delta_{23}\right]
\end{align*}
so 
\begin{align*}
F(q_{1},\dots,q_{4})=\delta_{12}\delta_{34}+\overline{\delta_{12}}\delta_{13}\delta_{24}+\overline{\delta_{12}}\overline{\delta_{13}}\delta_{14}\delta_{23}.
\end{align*}
Finally, if we express everything in terms of Kronecker deltas (using
$\overline{\delta_{ij}}:=1-\delta_{ij}$), we get 
\begin{align}
F(q_{1},\dots,q_{4})= & \delta_{12}\delta_{34}+\delta_{13}\delta_{24}+\delta_{14}\delta_{23}\nonumber \\
- & \delta_{12}\delta_{13}\delta_{24}-\delta_{12}\delta_{14}\delta_{23}-\delta_{13}\delta_{14}\delta_{23}\nonumber \\
+ & \delta_{12}\delta_{13}\delta_{14}\delta_{23}.\label{eq:F_deltas}
\end{align}
The left-hand side of the above equation can be divided in three groups
of terms: 
\begin{enumerate}
\item The first three terms correspond to all the pairwise contractions
of momenta, which gives a contribution of the form: 
\begin{align*}
3\left(\sum_{q}V^{2}(q)\right)^{2}=3\left(\overline{\epsilon^{2}}\right)^{2}=3\sigma_{\varepsilon}^{4};
\end{align*}
\item The following three involve double contractions (coincidences of momenta)
which imply $V(q_{1})=V(q_{2})=V(q_{3})=V(q_{4})$. This contribution
is $-3\sum_{q}V^{4}(q);$ 
\item The last term gives no contribution, since it implies that $q_{1}=-q_{2}=-q_{3}=-q_{4}$
and $q_{2}=-q_{3}$ . This will always yield a factor of $V(0)=0$. 
\end{enumerate}
Consequently, the four momentum sums of Eq.$\;$\ref{eq:V_sums} have
the value 
\begin{align}
3\left(\sum_{q}V^{2}(q)\right)^{2}-3\sum_{q}V^{4}(q)\label{eq:4th cumul}
\end{align}

This procedure is trivially generalized to any number of phase factors,
although the structure becomes rather complicated for higher order
terms. Fortunately, we will see that in certain limits, we may ignore
the contributions coming from the coincidences of momenta, and only
the pairwise contractions will contribute.

\subsection*{Phase Averages in the Gaussian Disorder Case}

In the case of the gaussian correlated disorder, the normalization
of the Fourier transform implies that $V^{2}(q)\sim\mathcal{O}(1/L)$.
The momentum sums give a factor of $\mathcal{O}(L)$, which means
that the two terms in Eq.$\;$\ref{eq:4th cumul} will be of order
\begin{align*}
3\left(\sum_{q}V^{2}(q)\right)^{2} & \sim\mathcal{O}(1),\\
3\left(\sum_{q}V^{4}(q)\right) & \sim\mathcal{O}(L)\times\mathcal{O}(\frac{1}{L^{2}})\sim\mathcal{O}(\frac{1}{L}).
\end{align*}
This means that the second term is negligible in the thermodynamic
limit. This argument can actually be carried through to any order,
since any term of the form $\sum_{q}V^{n}(q)$ goes to zero in the
limit $L\rightarrow\inf$, which renders all the contributions coming
from the coincidence of indices irrelevant in this limit.

Therefore, if we want to calculate a general $F(q_{1},\dots,q_{n})$,
we may only consider the sum of all pairwise contractions of momenta.
The total number of different contractions is $\left(n-1\right)!!$,
and each one contributes with a term $\left(\sum_{q}V^{2}(q)\right)^{\nicefrac{n}{2}}$to
the sum over momenta. Hence, we have 
\begin{align}
\sum_{q_{1}...q_{n-1}}V(q_{1})\;... & V(-q_{1}...-q_{n-1})\overline{e^{i\phi_{q_{1}}}...e^{i\phi_{-q_{1}...-q_{n-1}}}}=\nonumber \\
= & \left(n-1\right)!!\left(\sum_{q}V^{2}(q)\right)^{\frac{n}{2}}\left[1+\mathcal{O}\left(\frac{1}{L}\right)\right]\label{eq:F_q_gaussian}
\end{align}

\subsection*{Phase Averages in the Power-Law Disorder Case}

For the case of Power-Law Correlated Disorder, the Eq.$\;$\ref{eq:4th cumul}
is still valid, but one cannot generally ignore the $V^{4}$ term.
Let us consider only the cases where $\alpha>1$, meaning that 
\begin{equation}
V(q)=A(\alpha)\left(\frac{2\pi}{L}\right)^{\frac{1}{2}}\frac{1}{\left|q\right|^{\frac{\alpha}{2}}}
\end{equation}
with the normalization 
\begin{equation}
A(\alpha)=\frac{\sigma_{\varepsilon}}{\sqrt{2\mathcal{\zeta}(\alpha)}}\left(\frac{2\pi}{L}\right)^{\left(\alpha-1\right)/2}.\label{eq:Nomalize_const_power_law}
\end{equation}
Like before, we have $\sum_{q\neq0}V^{2}(q)=\sigma_{\varepsilon}^{2},$
but the calculation of $\sum_{q}V^{4}(q)$ is now, slightly different,
i.e. 
\begin{align*}
\sum_{q\neq0}V^{4}(q)= & A^{4}\left(\alpha\right)\left(\frac{2\pi}{L}\right)^{2}\sum_{q\neq0}\frac{1}{\left|q\right|^{2\alpha}}\\
= & 2A^{4}\left(\alpha\right)\left(\frac{2\pi}{L}\right)^{2\left(1-\alpha\right)}\sum_{p=1}^{L/2}\frac{1}{p^{2\alpha}}
\end{align*}
In the large $N$ limit, the last sum converges if $\alpha>1/2$ and
it gives $\zeta(2\alpha)$. Using Eq.~\ref{eq:Nomalize_const_power_law},
we finally obtain $\sum_{q\neq0}V^{4}(q)=\frac{\mathcal{\zeta}(2\alpha)}{2\mathcal{\zeta}(\alpha)}\sigma_{\varepsilon}^{4},$
which does not scale with the system size $L$. This interesting result
suggests that the argument made for the gaussian case does not work
here, and any calculation of the moments of $\rho(k,E)$ must account
for the coincidences of momenta. In fact, this is easily seen to be
true for any term of the form $\sum_{q}V^{2n}(q)$, yielding the general
form

\begin{equation}
\sum_{q}V^{2n}(q)=\frac{\mathcal{\zeta}(n\alpha)}{2^{n}\mathcal{\zeta}(\alpha)^{n}}\sigma_{\varepsilon}^{2n}\label{eq: v2n}
\end{equation}

Nevertheless, a special case happens when $\alpha\to1$. In this limit,
the denominator of Eq.$\;$\ref{eq: v2n} diverges as $(\alpha-1)^{-n}$
, while the numerator remains finite near $\alpha=1$. This means
that, for $\alpha\rightarrow1$ the corrections due to the coincidence
of momenta become negligible, and we have

\begin{align}
\sum_{q_{1}...q_{n-1}}V(q_{1})\;... & V(-q_{1}...-q_{n-1})\overline{e^{i\phi_{q_{1}}}...e^{i\phi_{-q_{1}...-q_{n-1}}}}=\nonumber \\
= & \left(n-1\right)!!\left(\sum_{q}V^{2}(q)\right)^{2n}\left[1+\mathcal{O}\left(\alpha-1\right)\right]\label{eq:F_q_powerlaw}
\end{align}
\global\long\def\newblock{}
  \bibliographystyle{plain}
\bibliography{CorrelatedDisorder-2.bib}

\end{document}